\documentclass[aps,prd,preprint,superscriptaddress,showpacs,floatfix,nobibnotes]{revtex4-1}

\usepackage{latexsym}
\usepackage{amsmath}
\usepackage{amssymb}
\usepackage{graphicx}
\usepackage{longtable}

\usepackage{bm}
\usepackage{epsfig}
\usepackage{subfigure}
\newcommand{\bea}{\begin{eqnarray}}
\newcommand{\eea}{\end{eqnarray}}
\newcommand{\xx}{\noindent}

\begin{document}

\title{Renormalization group methods and the 2PI effective action}

\author{M.E. Carrington}
\email[]{carrington@brandonu.ca} \affiliation{Department of Physics, Brandon University, Brandon, Manitoba, R7A 6A9 Canada}\affiliation{Winnipeg Institute for Theoretical Physics, Winnipeg, Manitoba}

\author{Wei-Jie Fu}
\email[]{w.fu@thphys.uni-heidelberg.de} \affiliation{Institut f\"{u}r Theoretische Physik, Universit\"{a}t Heidelberg,
Philosophenweg 16, 69120 Heidelberg, Germany
}

\author{D. Pickering}
\email[]{pickering@brandonu.ca} \affiliation{Department of Mathematics, Brandon University, Brandon, Manitoba, R7A 6A9 Canada}

\author{J.W. Pulver}
\email[]{joe89smith@hotmail.com} \affiliation{Department of Physics, Brandon University, Brandon, Manitoba, R7A 6A9 Canada}

\date{\today}

\begin{abstract}
We consider a symmetric scalar theory with quartic coupling in 4-dimensions
and compare the standard 2PI calculation with a modified version which uses an exact renormalization group method. 
The set of integral differential equations that are obtained from the exact renormalization group method truncate naturally, without the introduction of additional approximations. 
The results of the two methods agree well, which shows that the exact renormalization group can be used at the level of the 2PI effective action to obtain finite results without the use of counter-terms. 
The method therefore offers a promising starting point to study the renormalization of higher order $n$PI theories. 
\end{abstract}


\normalsize
\maketitle

\normalsize

\section{Introduction}

There is much interest in the study of non-perturbative systems, which cannot be solved by expanding in some small parameter. Two formalisms that have been proposed to address non-perturbative problems are $n$-particle irreducible ($n$PI) effective theories \cite{Jackiw1974,Norton1975}, and  
the exact renormalization group (RG) \cite{Wetterich1993,Ellwanger1993,Tetradis1994,Morris1994}.  

The 2PI formalism has been used to study finite temperature systems (see for example \cite{Blaizot1999,Berges2005a}),  
non-equilibrium dynamics and subsequent late-time thermalization (see \cite{Berges2001} and references therein), and transport coefficients \cite{Aarts2004,Carrington2006}. 
Beyond 2-loop order the 2PI effective action is not complete \cite{Berges-hierarchy} and one must use a higher order $n$PI theory. 
The 4PI effective action for scalar field theories is derived in Ref. \cite{Norton1975} using Legendre transformations. The method of successive Legendre transforms is used in \cite{Berges-hierarchy,Carrington2004}. A new method has been
developed to calculate the 5-loop 5PI and 6-loop 6PI effective action for scalar field theories \cite{Guo2011,Guo2012}.
The 3PI and 4PI effective actions have been used to obtain the integral equations from which
the leading order and next-to-leading order contributions to conductivity and shear viscosity can be calculated \cite{Carrington2009}.
However, the integral equations that are produced by higher order theories are difficult to solve, and new methods must be developed. 

One problem is the size of the phase space that is involved, but this can be significantly reduced using symmetry constraints \cite{Fu2013,Fu2014}. 
Another problem is the renormalization of higher order theories in 4-dimensions. 
The renormalization of the 2PI effective theory was only understood through the labours of multiple authors over a period of many years \cite{vanHees2002,Blaizot2003,Serreau2005,Serreau2010}. 
Using a diagrammatic approach, it was shown that renormalization requires introducing a set of vertex counter-terms, which obey different renormalization conditions and approach each other in the limit that the order of the approximation is taken to infinity. 
For higher order theories, the integral equations are too complicated to analyse in this way. 

The exact renormalization group has been applied to a variety of problems (for reviews see \cite{Morris-98,Bagnuls2001,Berges2002,Pawlowski2005,Rosten2007}) and has lead to insight into the nature of 
renormalizability. 
A regulator function is introduced which depends on the continuous parameter $\kappa$ whose role is to suppress fluctuations with momenta $q\le\kappa$ while leaving those with momentum $q > \kappa$ unaffected. The regulated action is equal to the standard action when $\kappa$ is taken to zero, which corresponds to removing the cutoff and including all fluctuations. On the other hand, when $\kappa\to\infty$ we can associate the regulated action with the classical action. The regulated action thus interpolates between the classical action and the full quantum action, as the parameter $\kappa$ is lowered to zero. The exact RG equations describe the evolution of the system from the scale of large $\kappa$, where the solutions are known, to the scale $\kappa=0$, where the solutions are desired. As always, physical quantities should be independent of the 
regularization scheme, which means in this case independent of the choice of regulator function. 

The formal relationship between the RG method and 2PI theories has been studied in \cite{Dupuis1,Blaizot-2PIa,Blaizot-2PIb,Dupuis2}, and the connection with higher $n$PI theories was developed in \cite{Carrington-BS}.
One of the difficulties with the standard RG flow equations obtained from the 1PI effective action is that they take the form of an infinite coupled hierarchy of functional differential equations, and an additional approximation is needed to truncate this hierarchy \cite{BMW1,BMW2}. 
An $n$PI effective theory also produces a infinite hierarchy of coupled integral equations, but in this case the hierarchy truncates automatically when the effective action is truncated at some order in the approximation (for example, a loop or $1/N$ expansion), and the truncation respects gauge invariance, to the order of the approximation \cite{Smit2003,Zaraket2004}. 
One advantage of the method we develop in this paper is that the RG flow equations that are obtained from the 2PI effective action also truncate naturally, and therefore do not require the introduction of additional approximations. 

In this paper we look at a specific 2PI calculation and show that it can be done in a different way, using a RG method. 
We start by calculating the 2-point and 4-point functions using the standard 2PI method, which was done previously in Ref. \cite{Berges2005a}. Then we do the calculation in a different way, without introducing counter-terms, using a regulated 2PI effective action and solving the resulting flow equations. The results of the two methods agree well, which shows that the RG method can be used at the level of the 2PI effective action to obtain finite results without the use of counter-terms. 

This paper is organized as follows. 

In section \ref{BSsection} we present our version of the 2PI calculation. 
In sections \ref{2pi-action} and \ref{2pi-n-point} we define the action and $n$-point functions.  The renormalizability of the 2PI theory is discussed in section \ref{renorm-2PI-section}. In \ref{2PI-numerical-section} we explain our numerical method, much of which will also be used in the RG calculation, which is presented in section \ref{RG-section}. In section \ref{FRG-formalism-section} we discuss the RG formalism, and in the following two sections, \ref{flow:action} and \ref{2PI-hiarchy}, we define the regulated action and obtain the general flow equations. In section \ref{flow-specific} we give the specific form of the flow equations when the effective action is truncated at order $\lambda^2$, and in section \ref{boundary-conditions-section} we derive the boundary conditions on the flow. In section \ref{truncation-section} we explain how the RG equations truncate. In section \ref{2pi-connection} we discuss the connection between the RG calculation and the standard 2PI one. Some details of the numerical method are given in \ref{RG-numerical-section}. In section \ref{results-all} we present our numerical results from both calculations. In section \ref{conclusions-section} we compare the two methods from the point of view of  computational difficulty, and present our conclusions. 

In most equations in this paper we suppress the arguments that denote the space-time dependence of functions. As an example of this notation, the quadratic term in the action is written:
\bea
\frac{i}{2}\int d^4 x\,d^4 y\,\varphi(x)G_{\rm no\cdot int}^{-1}(x-y)\varphi(y) ~~\longrightarrow~~\frac{i}{2}\varphi\, G_{\rm no\cdot int}^{-1}\varphi\,.
\eea
We use the notation $G_{\rm no\cdot int}$ for the bare propagator because we reserve $G_0$ for the propagator in the RG calculation in the limit that the regulator goes to zero. 


\section{The 2PI Effective Theory}
\label{BSsection} \vspace{5pt}

\subsection{2PI effective Action}
\label{2pi-action}

The classical action is 
\begin{eqnarray}
\label{action}
&& S[\varphi] =\frac{i}{2}\varphi \,G_{\rm no\cdot int}^{-1}\varphi -\frac{i}{4!}\lambda\varphi^{4}\,,~~
iG_{\rm no\cdot int}^{-1} = -(\Box + m^2)\,.
\eea
For notational convenience we use a scaled version of the physical coupling constant
($\lambda_{\,{\rm phys}} = i\lambda$). The extra factor of $i$ will be removed when rotating to Eucledian space to do numerical calculations. 
We consider the 2PI theory and construct the generating functional using 1- and 2-point sources
\begin{eqnarray}
\label{ZandW}
Z[J,J_{2}]&=&\int[d\varphi]\exp\bigg\{i\Big(S[\varphi]+J\varphi
+\frac{1}{2}J_{2}\varphi^2\Big)\bigg\}\,,\\[2mm]
W[J,J_{2}]&=&-i\ln Z[J,J_{2}]\,.\nonumber
\end{eqnarray}
Taking functional derivatives with respect to sources we 
obtain
\begin{eqnarray}
\label{expt-values}
\frac{\delta W[J,J_2]}{\delta J}
&=&\langle \varphi\rangle\equiv\phi\,,\\
\frac{\delta W[J,J_2]}{\delta
J_{2}}&=&\frac{1}{2}\langle\varphi^2\rangle = \frac{1}{2}(\phi^2 + G)\,.\nonumber
\end{eqnarray}
The 2PI effective action is obtained by taking the double Legendre transform of the generating functional $W[J,J_2]$ with respect to the sources $J$ and $J_2$ and taking $\phi$ and $G$ as the independent variables:
\begin{eqnarray}
\label{legendre-transform}
\Gamma[\phi,G]&=&W-J\frac{\delta W}{\delta J} -J_{2}\frac{\delta W}{\delta J_{2}} = W-J\phi-\frac{1}{2}J_{2}(\phi\phi+G)\,.
\end{eqnarray}
We write the result as a function of renormalized variables without introducing additional subscripts:
\begin{eqnarray}
\label{gamma-renorm}
&& \Gamma[\phi,G] =\Gamma_{\rm no\cdot int}[\phi,G] +\Gamma_{\mathrm{int}}[\phi,G]\,, \\[2mm]
&& \Gamma_{\rm no\cdot int}[\phi,G] = \frac{i}{2}\phi \,G_{\rm no\cdot int}^{-1}\phi+\frac{i}{2}\mathrm{Tr}\ln
G^{-1}
+\frac{i}{2}\mathrm{Tr}G_{\rm no\cdot int}^{-1}G\,, \nonumber\\[2mm]
&& \Gamma_{\mathrm{int}}[\phi,G] = \frac{i}{2}\phi \,\delta
G_{\rm no\cdot int}^{-1}\phi +\frac{i}{2}\mathrm{Tr}\delta
G_{\rm no\cdot int}^{-1}G-\frac{i}{4!}(\lambda+\delta\lambda)\phi^{4}-\frac{i}{4}(\lambda+\delta\lambda)\phi^{2}G+\Gamma_{2}[\phi,G;\lambda+\delta\lambda]\,,\nonumber\\[2mm]
&& i\delta G_{\rm no\cdot int}^{-1} = -(\delta Z \Box +\delta m^2)\,,~~~
iG_{\rm no\cdot int}^{-1} = -(\Box + m^2)\,, \nonumber 
 \end{eqnarray}
where $\Gamma_{2}$ contains all 2PI vacuum graphs whose vertices
are given by the terms cubic or quartic in $\varphi$ in the
expanded expression of $S[\phi+\varphi]-S[\phi]$.
%
Throughout this paper we use the notation $i\Gamma = \Phi$  where both $\Gamma$ and $\Phi$ carry the same subscripts or superscripts. For example, for the 2PI effective action we write $i\Gamma[\phi,G] = \Phi[\phi,G]$, for the 
interacting part of the 2PI effective action we have $i\Gamma_{\rm int}[\phi,G] = \Phi_{\rm int}[\phi,G]$, etc. 

The stationary condition is 
\bea
\label{stat-cond}
\frac{\delta \Phi[\phi,G]}{\delta G}\bigg|_{G=\tilde G} = 0\,.
\eea
The solution  $\tilde G = G(\phi)$ is an implicit function of the field. 
We define the resummed action
\bea
\label{resummed-action}
\tilde \Phi[\phi] = \Phi[\phi,G(\phi)]\,.
\eea
The minimum of the resummed action ($\tilde\phi$) satisfies the condition
\bea
\label{stat-cond2}
\frac{\delta\tilde \Phi[\phi]}{\delta\phi}\bigg|_{\tilde\phi} = 0\,.
\eea
In this paper we consider only the symmetric theory, which means we take $\tilde\phi=0$. 

\subsection{2PI $n$-point functions}
\label{2pi-n-point}

We can obtain $n$-point functions which obey the symmetries of the original theory by taking functional derivatives of the resummed action 
\bea
\label{resummed-npoint-generic}
\tilde \Phi^{(n)}[\phi] = \frac{\delta^n}{\delta \phi^n}\tilde \Phi_{\rm int}[\phi]\,.
\eea
These resummed $n$-point functions obey integral equations with kernels of the form
\bea
\label{kernels}
\Phi^{(n,m)}[\phi,G(\phi)] = 2^m\frac{\delta^{n+m}}{\delta\phi^n\delta G^m}\Phi_{\rm int}[\phi,G]\bigg|_{G=\tilde G}\,.
\eea
We introduce specific names for the kernels we will need:
\bea
\label{kernels-specific}
&& \Phi_\mathrm{int}^{(0,1)} = \Sigma^{(0,1)}\,,~~~\Phi_\mathrm{int}^{(2,0)} = \Sigma^{(2,0)}\,,\\
&& \Phi_\mathrm{int}^{(0,2)} = \Lambda^{(0,2)}\,,~~~\Phi_\mathrm{int}^{(2,1)} = \Lambda^{(2,1)}\,,~~~\Phi_\mathrm{int}^{(4,0)} = \Lambda^{(4,0)}\,.\nonumber
\eea
Both of the kernels denoted $\Sigma$ correspond to 2-point functions, and the kernels $\Lambda$ are 4-point functions. 
Using this notation the stationary condition in Eq. (\ref{stat-cond}) can be written:
\bea
\label{dyson1}
G^{-1}(\phi) = G_{\rm no\cdot int}^{-1}-\Sigma^{(0,1)}[\phi,G(\phi)]\,.
\eea
Thus we have a self-consistent equation for the propagator $G(\phi)$ which has the form of a Dyson equation.

From equations (\ref{resummed-action}, \ref{resummed-npoint-generic}, \ref{kernels}) we obtain integral equations for the resummed $n$-point functions
\bea
\label{resummed-npoint}
&& -\tilde\Phi^{(2)}[\phi] = G_{\rm no\cdot int}^{-1}-\Sigma^{(2,0)}[\phi,G(\phi)]\,,\\[2mm]
&& \tilde\Phi^{(4)}[\phi] = \Lambda^{(4,0)}[\phi,G(\phi)] + \frac{1}{2}\bigg(\Lambda^{(2,1)}[\phi,G(\phi)]~ G^2(\phi) ~\frac{\delta^2 \Sigma^{(0,1)}[\phi,G(\phi)]}{\delta \phi^2}~ + ~2~{\rm perms}\bigg)\,.
\eea
In both of these expressions we have dropped terms that contain kernels with an odd number of $\phi$ derivatives, since they will be zero in the symmetric theory. In addition, some terms have been dropped using the stationary condition (\ref{stat-cond}). 
The first equation looks like (\ref{dyson1}), but for an arbitrary truncation $\Sigma^{(0,1)}[\phi,G(\phi)] \ne \Sigma^{(2,0)}[\phi,G(\phi)]$ and therefore $-G^{-1}(\phi)\ne \tilde\Phi^{(2)}(\phi)$. 
In the second equation, the three terms in the round bracket represent the $s$, $t$ and $u$ channels of the 1-loop seagull diagram with one vertex $\Lambda^{(2,1)}$ and the other vertex given by:
\bea
M^\prime[\phi,G(\phi)] = \frac{\delta\Sigma^{(0,1)}[\phi,G(\phi)]}{\delta \phi^2}\,.
\eea
Using the chain rule and the kernel definitions in (\ref{kernels}, \ref{kernels-specific}) it is easy to show that the vertex $M^\prime$ satisfies the integral equation:
\bea
\label{Mprime-defn}
M^\prime[\phi,G(\phi)] = \Lambda^{(2,1)}[\phi,G(\phi)] + \frac{1}{2}M^\prime[\phi,G(\phi)]\; G^2(\phi)\, \Lambda^{(0,2)}[\phi,G(\phi)]\,,
\eea
where again we have dropped terms that are zero in the symmetric theory. 
This equation can be rewritten 
\bea
\label{Mprime2-defn}
&& M^\prime[\phi,G(\phi)] = \Lambda^{(2,1)}[\phi,G(\phi)] + \frac{1}{2} \Lambda^{(2,1)}[\phi,G(\phi)]\; G^2(\phi) M[\phi,G(\phi)] \,,
\eea
where the vertex $M$ is defined as
\bea
\label{M-defn}
&& M[\phi,G(\phi)] = \Lambda^{(0,2)}[\phi,G(\phi)] + \frac{1}{2}  \Lambda^{(0,2)}[\phi,G(\phi)]\; G^2(\phi)\; M[\phi,G(\phi)] \,.~~
\eea

In summary we have defined the following vertices:

\xx (1) Resummed vertices $\tilde\Phi^{(2)}$ and $\tilde\Phi^{(4)}$ ~~~~(\ref{resummed-npoint})

\xx (2) Kernels $\Sigma^{(2,0)}$, $\Sigma^{(0,1)}$, $\Lambda^{(4,0)}$, $\Lambda^{(2,1)}$ and $\Lambda^{(0,2)}$ ~~~~(\ref{kernels}, \ref{kernels-specific})

\xx (3) Bethe-Salpeter (BS) vertices $M^\prime$ and $M$~~~~ (\ref{Mprime-defn}, \ref{M-defn})

\noindent In the exact (untruncated) theory 
\bea
\label{exact2}
&& \Sigma_{\rm exact}^{(2,0)} = \Sigma_{\rm exact}^{(0,1)}~~~({\rm or} -\tilde\Phi_{\rm exact}^{(2)} = G_{\rm exact}^{-1})\,,\\[2mm]
\label{exact4}
&& \tilde\Phi_{\rm exact}^{(4)}=M_{\rm exact}=M_{\rm exact}^\prime\,.
\eea

We now impose the stationarity condition (\ref{stat-cond2}) and set $\phi=\tilde\phi=0$. We Fourier transform to momentum space and write the vertices as functions of their momentum arguments. (We do not introduce new notation to indicate that the function changes. For example, we should write $\Sigma^{(2,0)}[0,G(0)] \to \bar\Sigma^{(2,0)}(P)$), but we suppress the bar.) The vertices in the symmetric theory and in momentum space are written: 
\bea
\label{vertex-list}
&& {\rm resummed~vertices:}~~~~ \tilde\Phi^{(2)}(P)\,,~~\tilde\Phi^{(4)}(P,K,S)\,,~~\\
&& {\rm kernels:}~~~~ \Sigma^{(2,0)}(P)\,,~~\Sigma^{(0,1)}(P)\,,~~\Lambda^{(4,0)}(P,K,S)\,,~~\Lambda^{(2,1)}(P,K)\,,~~\Lambda^{(0,2)}(P,K)\,,\nonumber\\
&& {\rm BS~vertices:}~~~~ M^\prime(P,K)\,,~~M(P,K)\,.\nonumber
\eea

\subsection{2PI renormalization }
\label{renorm-2PI-section}

The 1PI effective action is renormalized by introducing three counter-terms in the Lagrangian (denoted $\delta Z$, $\delta m^2$ and $\delta\lambda$) which modify the bare parameters of the original theory (and the wave function normalization) and are determined by three renormalization conditions. To renormalize the 2PI theory we need multiple counter-terms, which we denote $\delta Z_i$, $\delta m_i^2$ and $\delta\lambda_i$. 
Counter-terms differentiated by different subscripts come from the same term in the Lagrangian, but correspond to different orders in the approximation that is used to truncate the effective action.
All counter-terms are determined from only three renormalization conditions. 
We describe the procedure below. 

One starts by adding counter-terms to each local, mass dimension 4 operator in the effective action
\bea
\label{L-Delta}
\Phi_\Delta = \frac{1}{4!} \Delta\lambda_4 \phi^{4}+\frac{1}{4}\Delta\lambda_{\rm tp} \phi^{2}G  +\frac{1}{8} \Delta\lambda_{\rm et} G^2\,.
\eea
In addition, one includes the usual counter-terms in the skeleton expansion of the effective action, to the approximation order. For example, to order $\lambda^3$ we have (see Fig. \ref{ct-fig}):
\bea
\label{L-ct}
\Phi_{\rm ct}&& = -\frac{i}{2}(\delta Z_2\Box+\delta m_2^2)\phi^2
-\frac{i}{2}(\delta Z_0\Box+\delta m_0^2) {\rm Tr}G
+ \frac{1}{4!} \delta\lambda^\prime_4 \phi^{4}
+ \frac{1}{4}\delta\lambda^\prime_{\rm tp} \phi^{2}G \nonumber\\
&& +\frac{1}{3}\lambda\,\delta\lambda_{\rm egg}\phi^2 G^3
 +\frac{1}{8} \delta\lambda^\prime_{\rm et} G^2  
+\frac{1}{24} \delta\lambda_{\rm bb} \lambda G^3 +{\cal O}(\lambda^4)\,.
\eea
\begin{figure}
\begin{center}
\includegraphics[width=14cm]{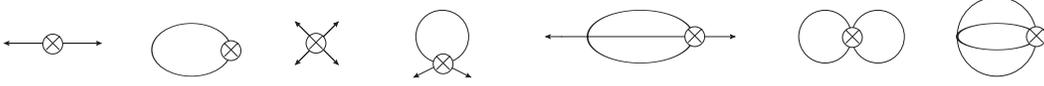}
\end{center}
\caption{Contributions to $\Phi_{\rm ct}$ to order $\lambda^3$. The diagrams represent the terms in Eq. (\ref{L-ct}) in the order they appear in the equation.  \label{ct-fig}}
\end{figure}
Primes are used for counter-terms which have partners in (\ref{L-Delta}). We define
\bea
\label{lambda-sum}
\delta\lambda_4 =\delta\lambda^\prime_4+\Delta\lambda_4\,,~~~\delta\lambda_{\rm tp} =\delta\lambda^\prime_{\rm tp}+\Delta\lambda_{\rm tp}\,,~~~\delta\lambda_{\rm et} =\delta\lambda^\prime_{\rm et} + \Delta\lambda_{\rm et}\,.
\eea
The coupling counter-terms in $\Phi_{\rm ct}$ are chosen to cancel divergences in the integrals in the 4-kernels, and the coupling counter-terms in $\Phi_\Delta$ cancel the remaining divergences in the resummed 4-point vertices. The 4-kernels are divergent (for example $\Lambda^{(0,2)} = \Delta\lambda_{\rm et}+ \Lambda_f^{(0,2)}$ where $\Lambda_f^{(0,2)}$ is the quantity that is made finite by $\lambda_{\rm et}^\prime + \lambda_{\rm bb} + \cdots $), but this is not a problem since the 4-kernels are not directly related to physical quantities. 
The 2-point functions contain coupling constant counter-terms that must be obtained self-consistently from the appropriate BS equation. 
The vertices and their corresponding counter-terms are listed in Table \ref{table-ct}.
\begin{table}
\begin{center}
\begin{tabular}{|l|c|c|c |} 
\hline
 ~  & c-term \big| vertex & c-term \big| vertex & c-term \big| vertex \\\hline\hline
{\rm 4-pt~kernels}   
& ~$\delta\lambda_4^\prime +\cdots$~ \big| $\Lambda^{(4,0)}$ 
& ~ $\delta\lambda_{\rm tp}^\prime + \delta\lambda_{\rm egg} + \cdots$ ~\big| $\Lambda^{(2,1)}$ 
& ~  $\delta\lambda_{\rm et}^\prime + \delta\lambda_{\rm bb} +\cdots$~ \big| $\Lambda^{(0,2)}$ ~ \\\hline
{\rm 4-pt~fcns}  
&    $\Delta\lambda_4$ ~ \big| $\tilde\Phi^{(4)}$           
&    $\Delta\lambda_{\rm tp}$ \big| $M^\prime$     
& $\Delta\lambda_{\rm et}$  \big| $M$ ~  \\\hline
{\rm 2-pt~fcns}  
&    $\delta Z_2$, $\delta m^2_2$, $\delta\lambda_{\rm tp}$ ~ \big| $\tilde\Phi^{(2)}$ 
&    $\delta Z_0$, $\delta m^2_0$, $\delta\lambda_{\rm et}$ ~ \big| $G$           
&    
 ~  \\\hline
\end{tabular}
\end{center}
\caption{Counter-terms for the various vertices in the 2PI theory\label{table-ct}}
\end{table}

In the truncated theory, the different $n$-point functions in (\ref{vertex-list}) are not the same, and similarly the counter-terms which are differentiated by subscripts are not the same. Renormalizability requires only that the untruncated (exact) theory contains one mass counter-term, one wave-function renormalization counter-term and one coupling constant counter-term, which produce one renormalized $2$-point function (\ref{exact2}) and one renormalized 4-point function (\ref{exact4}).
The counter-terms introduced in (\ref{L-Delta} - \ref{lambda-sum}) must therefore satisfy $\{\delta m_2^2,\delta m_0^2\}\to\delta m^2$, $\{\delta Z_2,\delta Z_0\}\to\delta Z$ and $\{\delta\lambda_4,\delta\lambda_{\rm tp},\delta\lambda_{\rm egg},\delta\lambda_{\rm et},\delta\lambda_{\rm bb} \cdots\}\to\delta\lambda$ when the order of the approximation is taken to infinity. To obtain this result, the counter-terms which carry different subscripts 
must be determined from the same renormalization condition, which means that the two 2-point functions must satisfy the same  renormalization conditions, and the three 4-point functions must satisfy one renormalization condition
\bea
\label{renorm-conds1}
&& i\tilde\Phi^{(2)}(0) = -iG^{-1}(0) = m^2 \,,\\[2mm]
&& i\frac{d}{dP^2}\tilde\Phi^{(2)}\bigg|_{P=0} = -i\frac{d}{dP^2}G^{-1}\bigg|_{P=0} = -1 \,,\nonumber\\[2mm]
&& \tilde\Phi^{(4)}(0,0,0,0) = M(0,0) = M^\prime(0,0) = \lambda \,,\nonumber
\eea
where the notation $(0)$, $(0,0)$, etc., indicates that all momentum components of each leg are set to zero. 

\subsection{2PI Numerical Method}
\label{2PI-numerical-section}

In this paper we truncate the effective action so that it includes all terms of order $\lambda^2$ in the skeleton expansion. To this order $\Sigma^{(0,1)}=\Sigma^{(2,0)}$, $\Lambda^{(0,2)}=\Lambda^{(2,1)}$ and $M=M^\prime$, and therefore we consider only $\Sigma^{(0,1)}$, $\Lambda^{(0,2)}$ and $M$, and we drop the superscripts. 
The only counter-terms we will need are $\delta Z_0$, $\delta m^2_0$, $\delta \lambda_{\rm et}^\prime$ and $\Delta \lambda_{\rm et}^\prime$, and therefore we drop the counter-term subscripts as well. Furthermore, at order $\lambda^2$ the kernel $\Lambda$ does not contain any non-global divergences that require a coupling counter-term, which means that the division of the counter-term into two pieces is not necessary, and therefore we use only $\delta\lambda_{\rm et}\equiv\delta\lambda$.

Equations (\ref{kernels}, \ref{kernels-specific}, \ref{dyson1}, \ref{M-defn}) determine the self-energy, propagator, 4-kernel and BS vertex. We rotate to Eucledian space, discretize, and solve the resulting set of equations using an iterative relaxation method. These three steps are described in the following three sub-sections. More details can be found in Refs. \cite{Fu2013,Fu2014}

\subsubsection{Eucledian space equations}

Using the effective action in (\ref{gamma-renorm}) with $\Phi_2 = i\Gamma_2$ shown in Fig. \ref{phi-fig}, the kernel and self-energy in momentum space are:
\begin{eqnarray}
\label{Lam-mink}
&& \Lambda(P,K)=(\lambda+\delta\lambda)+\frac{\lambda^{2}}{2}\int dQ G(Q)G(Q+P-K) + \frac{\lambda^{2}}{2}\int dQ
G(Q)G(Q+P+K)\,, \\
&&  \Sigma(P) = i(\delta Z P^{2}-\delta
m^{2})+(\lambda+\delta\lambda)\frac{1}{2}\int dQ G(Q)  +\frac{\lambda^{2}}{6}\int dQ\int dL
G(Q)G(L)G(Q+L+P)\,.\nonumber\\
&& \label{sigma-mink}
\end{eqnarray}
where we have used $dQ = \frac{d^{4}q}{(2\pi)^{4}}$.
\begin{figure}
\begin{center}
\includegraphics[width=10cm]{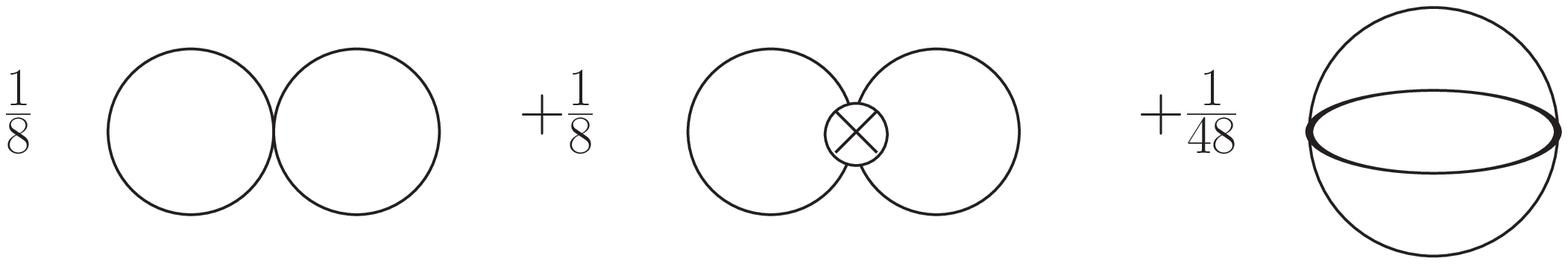}
\end{center}
\caption{Contributions to $\Phi_2 = i\Gamma_2$ to order $\lambda^2$.\label{phi-fig}}
\end{figure}
The propagator and the BS vertex are given by:
\bea
\label{prop-mink}
&& G^{-1} = G_{\rm no\cdot int}^{-1} -  \Sigma(Q)\,,\\
\label{M-mink}
&& M(P,K)=\Lambda(P,K)+\frac{1}{2}\int dQ\Lambda(P,Q)G^2(Q) M(P,K)\,.
\eea

We rotate to Eucledian space and define the Eucledian variables:
\bea
\label{euc-defn}
&& q_0 ~\to~ i q_4\,,~~dQ ~\to~i dQ_E\,,~~Q^2~\to ~ -Q_E^2 \,,\\
&& m^2 ~\to~ m^2_E\,,~~\delta m^2 ~\to~ \delta m^2_E\,,~~\delta Z ~\to~ \delta Z_E\,,\nonumber\\
&& \lambda ~\to~ -i \lambda_E
\,,~~\delta \lambda ~\to~ -i\delta \lambda_E\,,\nonumber\\
&& G^{-1}_{\rm no\cdot int}~\to ~ i (G^{-1}_{\rm no\cdot int})_E\,,~~\Sigma~\to~ -i \Sigma_E  ~~\Rightarrow~~     G^{-1}~\to ~ i G_E^{-1}  = i(P_E^2+m_E^2+\Sigma_E)\,,\nonumber \\
&& \Lambda ~\to~ i\Lambda_E\,,~~M ~\to~ i M_E\,.\nonumber
\eea
The extra factor of $i$ in the definition of the coupling removes the $i$ that was introduced in the definition $\lambda_{\rm phys} = i\lambda$. 
In Eucledian space equations (\ref{Lam-mink} - \ref{M-mink}) become:
\begin{eqnarray}
\label{Lam-euc}
&& \Lambda_E(P,K) = -(\lambda_E+\delta\lambda_E)\\
&&~~~~~~~~~~~~~ +\frac{\lambda_E^{2}}{2}\int dQ_E G_E(Q)G_E(Q+P-K) + \frac{\lambda_E^{2}}{2}\int dQ_E G_E(Q)G_E(Q+P+K)\,,\nonumber \\[2mm]
\label{sigma-euc}
&&  \Sigma_E(P) = \delta Z_E P^{2}+\delta m_E^{2} \\
&&~~~~~~~~~ + (\lambda_E+\delta\lambda_E)\frac{1}{2}\int dQ_E G_E(Q)  - \frac{\lambda_E^{2}}{6}\int dQ_E\int dL_E G_E(Q)G_E(L)G_E(Q+L+P)\,,\nonumber\\[2mm]
\label{prop-euc}
&& G_E^{-1} = (G_{\rm no\cdot int}^{-1})_E + \Sigma_E(Q) \,,\\[2mm]
\label{M-euc}
&& M_E(P,K)=\Lambda_E(P,K) +\frac{1}{2}\int dQ_E \,\Lambda_E(P,Q)G_E^2(Q) M_E(P,K) \,.
\eea
From now on we suppress the subscripts $E$ indicating Eucledian space. 
The counter-terms are determined from the renormalization conditions (\ref{renorm-conds1}) which are written in Eucledian space 
\bea
\label{renorm-conds2}
&& G^{-1}(0) =m^2\,,~~ \frac{d}{dP^2}G^{-1}\bigg|_{P=0} =1\,,~~  M(0) =- \lambda \,.
\eea


\subsubsection{Discretization}
\label{discretization-section}

In order to do the numerical calculation, we restrict to a box in co-ordinate space of finite volume $L^3 \beta$. Fourier transforming to momentum space one obtains discrete frequencies and momenta. This can be written
\bea
&& \int \frac{dp_4}{2\pi} \prod_{i=1}^3 \int^\infty_{-\infty}\frac{dp_i}{2\pi} f(p_4,p_i)
\to \frac{m_t m_s^3}{(2\pi)^4} \sum_{n_{4}=-\frac{N_t}{2}+1}^{\frac{N_t}{2}} \prod_{i=1}^3 \sum_{n_i=-\frac{N_s}{2}+1}^{\frac{N_s}{2}} f(m_t n_4,m_s n_i) \,,\\[4mm]
\label{mhat-defn}
&& m_t = 2\pi T = 2\pi/(N_t a_t)\,,~~m_s= 2\pi L^{-1} = 2\pi/(N_s a_s)\,,~~
L=a_s N_s\,,~~T=1/a_t N_t\,.
\eea
The parameters $a_t$ and $a_s$ are the lattice spacing in the temporal and spatial directions. Indices which fall outside of the range $\{-N/2+1,N/2\}$ are wrapped inside using periodic boundary conditions. This is done using the function
\bea
\label{reindex}
{\rm rndx}[{\rm index}] = 1 - N/2 + {\rm Mod}[{\rm index} + N/2 - 1, N]\,,
\eea
where Mod[$m$,$n$] is an integer function that gives the remainder on division of $m$ by $n$ so that $0<$ Mod[$m$,$n$] $<n-1$ (for example, Mod[17,17]=0 and Mod[23,17]=6).

After discretization the bare propagator has the form
\bea
G_{\rm no\cdot int}^{-1}(P) \to G_{\rm no\cdot int}^{-1}(m_t n_4,m_s n_1,m_s n_2,m_s n_3) = m_t^2 n_4^2+m_s \sum_{i=1}^3 n_i^2+m^2\,.
\eea
To simplify the notation we represent the arguments of a function of discrete variables using one boldface character, for example, 
\bea
&& G(P) \to G(m_t n_4,m_s n_1,m_s n_2,m_s n_3) \to G({\bf n})\,,\\
&& M(P,K) = M(m_t n_4,m_s n_1,m_s n_2,m_s n_3;m_t j_4,m_s j_1,m_s j_2,m_s j_3) \to M({\bf n},{\bf j})\,.\nonumber
\eea
We also write the four summations which correspond to one discretized 4-momentum integral as one summation, for example,
\bea
\frac{m_t m_s^3}{(2\pi)^4}\sum_{n_{4}=-\frac{N_t}{2}+1}^{\frac{N_t}{2}} \prod_{i=1}^3 \sum_{n_i=-\frac{N_s}{2}+1}^{\frac{N_s}{2}} = \sum_{\bf n}\,.\nonumber
\eea

The scalar $\phi^4$ theory in 4-dimensions is non-interacting if it is considered
as a fundamental theory valid for arbitrarily high momentum scales (quantum triviality), but the renormalized coupling is non-zero if the theory has an ultra-violet cutoff and an infra-red regulator. In our calculation the lattice spacing parameters $a_t$ and $a_s$ provide an ultra-violet cutoff for the $p_4$ and $p_i$ momentum integrals, and the mass $m$ regulates the momentum integrals in the infra-red.


The discretized forms of equations (\ref{Lam-euc} - \ref{M-euc}) are
\begin{eqnarray}
\label{prop-discrete}
&& G^{-1}({\bf n})=G_{\rm no\cdot int}^{-1}({\bf n})+\Sigma({\bf n})\,,\\
&& ~~~~~~ G_{\rm no\cdot int}^{-1}({\bf n}) = m_t^2 n_4^2+ m_s^2 \sum_{i=1}^3 n_i^2 + m^2\,,\nonumber\\
\label{sigma-discrete}
&& \Sigma({\bf n}) = \delta Z (m_t^2 n_4^2+ m_s^2 \sum_{i=1}^3 n_i^2)+\delta
m^{2} +\hat \Sigma({\bf n})\,,\\
\label{hatsigma-discrete}
&& ~~~~~~ \hat \Sigma({\bf n}) = (\lambda+\delta\lambda)\frac{1}{2} \,\sum_{\bf j}G({\bf j})
 -\frac{\lambda^{2}}{6} \,\sum_{\bf j}\sum_{\bf k}
G({\bf j})G({\bf k})G({\bf j}+{\bf k}+{\bf n})\,,\nonumber\\[6mm]
\label{M-discrete}
&&M({\bf n},{\bf 0}) = \Lambda({\bf n},{\bf 0})+ \frac{1}{2}  \,\sum_{\bf j}\Lambda({\bf n},{\bf j})
G^{2}({\bf j})M({\bf j},{\bf 0})\,,\\[6mm]
\label{Lam-discrete}
&& \Lambda({\bf n},{\bf l}) = -\delta\lambda + \hat\Lambda({\bf n},{\bf l}) \,,\\[2mm]
\label{hatLam-discrete}
&& ~~~~~~ \hat\Lambda({\bf n},{\bf l}) = -\lambda+ \frac{\lambda^{2}}{2} \,\sum_{\bf j}
G({\bf j})G({\bf j} + {\bf n} - {\bf l})  +\frac{\lambda^{2}}{2} \,\sum_{\bf j}
G({\bf j})G({\bf j} + {\bf n} + {\bf l})\,. 
\end{eqnarray}
Note that since the BS equation resums only one channel  one can fix the momentum on one side of the vertex $M$. 
The counter-terms are determined from the renormalization conditions (\ref{renorm-conds2}) which now take the form:
\bea
\label{rc-m}
&& G^{-1}(0,0,0,1)-G^{-1}({\bf 0}) = m_s^2 ~~\Rightarrow~~\delta Z = \big[\hat\Sigma({\bf 0})-\hat\Sigma(0,0,0,1)\big]\frac{1}{m_s^2}\,,\\
\label{rc-Z}
&& G^{-1}({\bf 0})=m^2  ~~\Rightarrow~~  \delta m^2 = -\hat\Sigma({\bf 0})\,, \\[2mm]
\label{rc-lam}
&& M({\bf 0},{\bf 0})=-\lambda \,. 
\eea

\subsubsection{The relaxation method}
\label{2PI-iterations-section}

We solve the system of equations (\ref{prop-discrete} - \ref{rc-lam}) using an iterative relaxation method. We use an index in round brackets to indicate the iteration number of a given quantity. 
In the first step of the calculation, the counter-terms are determined at fixed temperature $T_0\ll m$. We will verify numerically that $T_0$ corresponds to the zero temperature limit, and we refer to it from here on as zero temperature. 
We will study the temperature dependence of the $n$-point functions by decreasing $N_t$, using the counter-terms obtained at zero temperature. From this point on, we scale all dimensionful variables by $m$, or equivalently we set the renormalized mass to one and express all quantities in mass units. 

First we describe the general method to find the counter-terms by imposing the renormaliztion conditions at $T_0$. 
The zeroth iteration of the propagator is the bare propagator. 
The kernel $\Lambda$ at any iteration order is obtained from (\ref{Lam-discrete}, \ref{hatLam-discrete}) using propagators at the corresponding iteration order. The BS vertex $M$ at zeroth iteration order is defined to be $\Lambda$ at zeroth order. Thus we have
\bea
\label{31}
&& G^{(0)}({\bf n}) = \big(m_t^2 n_4^2+ m_s^2 (n_1^2+n_2^2+n_3^2)+1^2\big)^{-1} \,,\\[2mm]
\label{32}
&& \hat\Lambda^{(0)} = \hat\Lambda[G^{(0)}]\,,~~ \delta\lambda^{(0)} = \lambda+\hat\Lambda^{(0)}({\bf 0},{\bf 0})\,,~~M^{(0)} = -\delta\lambda^{(0)}+\hat\Lambda^{(0)}\,.
\eea
Starting from these zeroth order solutions we iterate to find self-consistent solutions 
\bea
\label{33}
&& \hat\Sigma^{(i+1)} = \hat\Sigma[G^{(i)},\delta\lambda^{(i)}]\,, \\
\label{34}
&& \delta Z^{(i+1)} = \big[\hat\Sigma^{(i+1)}(0,0,0,0)-\hat\Sigma^{(i+1)}(0,0,0,1)\big]\frac{1}{m_s^2}\,,\\
\label{35}
&& (\delta m^2)^{(i+1)} = -\hat\Sigma^{(i+1)}(0,0,0,0)\,,\\
\label{36}
&&(G^{(i+1)})^{-1} = G_0^{-1} + \hat\Sigma^{(i+1)} + \delta Z^{(i+1)} \big[m_t^2 j_4^2 + m_s^2 \sum_{i=1}^3 j_i^2\big] + (\delta
m^{2})^{(i+1)}\,, \\
\label{37}
&& \hat\Lambda^{(i+1)} = \hat\Lambda[G^{(i+1)}]\,,\\
&&M^{(i+1)}({\bf i},{\bf 0})=  \big[\hat\Lambda^{(i+1)}({\bf i},{\bf 0})-\delta\lambda^{(i)}\big]+ \frac{1}{2}  \sum_{\bf n}\big[\hat\Lambda^{(i+1)}({\bf i},{\bf n}) -\delta\lambda^{(i)}\big]
\big(G^{(i+1)}({\bf n})\big)^2M^{(i)}({\bf n},{\bf 0})\,,\nonumber\\
\label{39}\\
\label{38}
&&
\delta\lambda^{(i+1)} = \lambda+\delta\lambda^{(i)}+M^{(i+1)}({\bf 0},{\bf 0})\,.
\eea
Iterations are terminated when the relative maximum difference between the $(i+1)$th iteration and the $i$th, for any quantity, at any point in the phase space, is less than $10^{-4}$. 


If we simply ignore the vertex counter-term, the equations that determine the 2-point function are independent of those that determine the 4-vertex. This means that if renormalization were not necessary, one could calculate the propagator independently, and use this result in the BS equation to calculate the vertex $M$.
Once the counter-terms have been determined at zero temperature, calculations at different finite temperatures are much simpler because of the fact that when the counter-terms are known, the equations that give the 2-point function {\it are} decoupled from the vertex equations. This means that one can solve the equations at any finite temperature using a simpler procedure.  Symbolically:
\bea
\begin{array}{lll}
& G^{(0)} ~\to~ G^{(1)} ~\to ~ \cdots  G^{(i_{\rm final})}\equiv G &~~~~~~~{\rm using}~~~(\ref{prop-discrete}, \ref{sigma-discrete}) \\[4mm]
& \Lambda = \Lambda[G]&~~~~~~~{\rm using}~~~(\ref{Lam-discrete}, \ref{hatLam-discrete})  \\[4mm]
& M^{(0)} = \Lambda~~{\rm and}~~
M^{(0)} ~\to~ M^{(1)} ~\to ~ \cdots M^{(i_{\rm final})} \equiv M& ~~~~~~~{\rm using}~~~(\ref{M-discrete})
\end{array}\nonumber
\eea
If the number of iterations it takes to obtain convergence of the self-energy equation is $N_G$ and the number it takes to converge the BS equation is $N_M$, the first (zero temperature) procedure requires $N_G\times N_M$ iterations, and the second (finite temperature) calculation takes $N_G + N_M$ iterations. Typically $N_G\sim N_M\sim 5$ and therefore, after the renormalization is performed, subsequent calculations at different finite temperatures are much quicker. 

The coupling of the equations for the 2-point and 4-point functions that we have described above is a general feature of the zero temperature calculation at arbitrary approximation order. However, when the effective action is truncated at order $\lambda^2$ as in this paper, the zero temperature calculation can be done in a simpler way. The basic reason is that the vertex counter-term contributes to the self-energy only through the momentum independent tadpole diagram, and therefore one can proceed as follows.

\begin{enumerate}

\item set $\delta\lambda=0$ and drop the tadpole contribution in the self-energy in (\ref{sigma-discrete})

\item relax the Dyson equation (\ref{prop-discrete}) using the renormalization conditions (\ref{rc-m}, \ref{rc-Z}) to obtain $G_{\rm temp}({\bf j})$, $\delta m^2_{\rm temp}$ and $\delta Z$

\item use $G_{\rm temp}$ in (\ref{hatLam-discrete}) to get $\hat\Lambda_{\rm temp}({\bf n},{\bf l})$ and define $\delta\lambda^{(0)} =\lambda+ \hat\Lambda_{\rm temp}({\bf 0},{\bf 0})$

\item iterate the BS equation (\ref{M-discrete}) starting from $M=-\delta\lambda^{(0)} + \hat\Lambda_{\rm temp}$ and obtain $\delta\lambda$

\item calculate the tadpole term in the self-energy (\ref{sigma-discrete}) using $G_{\rm temp}$ and $\delta\lambda$

\item use the renormalization condition (\ref{rc-m}) to get $(\delta m^2)^\prime$

\item the full mass counter-term is the sum $\delta m^2=\delta m^2_{\rm temp}+(\delta m^2)^\prime$
\end{enumerate}
The total number of steps is $N_G + N_M +1$. 

\vspace*{.2cm}

We will do the numerical calculations using the renormalized parameters $m=1$ and $\lambda=1$. The renormalization is done with $N_t=128$, $N_s=32$, $a_t=1/16$ and $L=a_s N_s=2$, and finite temperature calculations are done with $126\ge N_t\ge 8$. We present our results in section \ref{results-all}, together with the results from the RG calculation, which is described in the next section.  All calculations are done using fast Fourier transforms, to improve performance. 
\newpage

\section{Exact Renormalization Group Calculation}
\label{RG-section}

\subsection{The RG formalism}
\label{FRG-formalism-section}

Using the functional renormalization group method, 
we add to the action in (\ref{action}) a non-local regulator term
\begin{eqnarray}
\label{action-RG}
S_{\kappa}[\varphi]=S[\varphi]+\Delta S_{\kappa}[\varphi]\,,~~~
\Delta  S_{\kappa}[\varphi] = -\frac{1}{2}\hat R_{\kappa}\varphi^2\,. 
\eea
The bare mass and coupling are defined at an ultra-violet scale $\mu$ which must be specified (we use $\mu$ instead of the traditional $\Lambda$ because that letter has already been used for the 4-point kernels).
The parameter $\kappa$ has dimensions of momentum and the regulator $\hat R_{\kappa}(Q)$ is chosen to have the following
properties: when $Q\ll \kappa$, $\hat R_{\kappa}(Q)\sim \kappa^{2}$, and when $Q\geq\kappa$, $\hat R_{\kappa}(Q)\rightarrow 0$. The effect is therefore that (1) for $Q\ll \kappa$ the regulator is a large mass term 
which suppresses quantum fluctuations with wavelengths $1/Q\gg
1/\kappa$ and; (2) fluctuations with $Q\gg\kappa$ and wavelengths $1/Q\ll
1/\kappa$ are not affected by the presence of the regulator.

The $n$-point functions of the theory depend on the parameter $\kappa$ and the goal is to calculate them in the limit $\kappa\to 0$, where the full quantum theory is restored.
One obtains a hierarchy of coupled differential `flow' equations for the derivatives of the $n$-point functions with respect to $\kappa$. We will show that when the 2PI effective action is used, this hierarchy is truncated when the effective action is. 
One chooses $\mu$ large enough that when $\kappa=\mu$ the theory is classical and the 2- and 4-point functions are known functions of the bare parameters. 
The flow equations can then be integrated from the scale $\kappa=\mu$, using the known classical solutions as boundary conditions, to the scale $\kappa=0$, at which the desired quantum solutions are obtained. 

A fundamental technical difficulty with the RG formalism is created by the fact that the renormalization conditions 
are defined in terms of the quantum ($\kappa\to 0$) $n$-point functions, which are obtained only after the calculation is finished.  
In the 2PI calculation described in the previous section, we choose values for the renormalized mass and coupling,  and input them into the calculation (the renormalized mass defines our system of units and thus we always choose $m=1$). 
Using the RG method, we want to specify chosen values for the renormalized mass and coupling, as before, but the required calculational input is the bare parameters, not the renormalized ones. The {\it result} of the calculation is the momentum dependent quantum $n$-point functions, which give (at zero momentum) the values of the renormalized mass and coupling. 
An arbitrary choice of the bare parameters will not produce the chosen renormalized parameters, and we do not know in advance which choice of bare parameters will. We must ``tune" the bare parameters, so that the renormalized mass and coupling that are produced by the calculation are the ones that were orginally specified. 

A summary of the procedure is:

\noindent (1) chose goal values for the renormalized mass and coupling

\noindent  (2) start with a ``guess'' for the corresponding values of the bare parameters defined at $\kappa=\mu$

\noindent (3) integrate the flow equations starting from classical solutions which are functions of the bare parameters and obtain the $n$-point functions at the scale $\kappa=0$

\noindent (4) extract the corresponding renormalized parameters and compare with the goal values

\noindent (5) adjust the bare parameters up or down accordingly and return to step (3)

\noindent One repeats steps (3) - (5) until the bare parameters are found that produce the desired renormalized ones. 
The end result is momentum dependent quantum $n$-point functions which satisfy the chosen renormalization conditions. 

As is the case for the 2PI calculation, the finite temperature calculation is simpler than the zero temperature one. We increase the temperature by decreasing $N_t$, and integrate the flow equations from $\kappa=\mu$ to $\kappa=0$, starting from the classical solutions and using the bare parameters obtained from the zero temperature calculation. 

\subsection{The 2PI FRG effective action}
\label{flow:action}

The 2PI effective action in  the FRG formalism is obtained from equations (\ref{ZandW}, \ref{expt-values}) using the regulated action (\ref{action-RG}): 
\begin{eqnarray}
\label{ZandW-RG}
&& Z_\kappa[J,J_{2}] = \int[d\varphi]\exp\bigg\{i\Big(S[\varphi]+J\varphi
+\frac{1}{2}J_{2}\varphi^2 - \frac{1}{2}\hat R_\kappa\varphi^2\Big)\bigg\}\,,\\[2mm]
\label{ZandW-RG-2}
&& W_\kappa[J,J_{2}] = -i\ln Z_\kappa[J,J_{2}]\,,\\[2mm]
\label{expt-values-RG}
&& \frac{\delta W_\kappa[J,J_2]}{\delta J}
 = \langle \varphi\rangle\equiv\phi\,,~~~ \frac{\delta W_\kappa[J,J_2]}{\delta
J_{2}} = \frac{1}{2}\langle\varphi^2\rangle = \frac{1}{2}(\phi^2 + G)\,.
\end{eqnarray}
The expectation values are calculated in the presence of the regulator and therefore depend on the parameter $\kappa$, which means that the relations between $(\phi,G)$ and $(J,J_2)$ are $\kappa$-dependent.
The 2PI effective action is obtained by taking the double Legendre transform of the generating functional $W_\kappa[J,J_2]$ with respect to the sources $J$ and $J_2$ and taking $\phi$ and $G$ as the independent variables (see equation (\ref{legendre-transform})):
\begin{eqnarray}
\label{legendre-transform-RG}
\hat\Gamma_\kappa[\phi,G]&=&W_\kappa-J\frac{\delta W_\kappa}{\delta J} -J_{2}\frac{\delta W_\kappa}{\delta J_{2}} = W_\kappa-J\phi-\frac{1}{2}J_{2}(\phi\phi+G)\,.
\end{eqnarray}
After performing the Legendre transform, the functional arguments of the effective action $\phi$ and $G$ are independent of the regulator function and the parameter $\kappa$, but the non-interacting propagator does depend on $\kappa$. 
We define 
\bea
\label{G0-first}
iG_{\mathrm{no}\cdot \mathrm{int}\cdot\kappa}^{-1} = iG_{\mathrm{no}\cdot \mathrm{int}}^{-1}-\hat R_\kappa = -\Box-(m^2+\hat R_k) \,.
 \end{eqnarray}
Using this notation the effective action $\hat\Gamma_\kappa[\phi,G]$ can be written (see equation (\ref{gamma-renorm}))
\begin{eqnarray}
\label{gamma-RG}
&& \hat\Gamma_\kappa[\phi,G] =\Gamma_{\mathrm{no}\cdot \mathrm{int}\cdot\kappa}[\phi,G] +\Gamma_{\mathrm{int}}[\phi,G]\,, \\[2mm]
&& \hat\Gamma_{\mathrm{no}\cdot \mathrm{int}\kappa}[\phi,G] = \frac{i}{2}\phi \,G_{\mathrm{no}\cdot \mathrm{int}\cdot\kappa}^{-1}\phi+\frac{i}{2}\mathrm{Tr}\ln
G^{-1}
+\frac{i}{2}\mathrm{Tr}G_{\mathrm{no}\cdot \mathrm{int}\cdot\kappa}^{-1}G \,,\nonumber\\[2mm]
&& \Gamma_{\mathrm{int}}[\phi,G] =  -\frac{i}{4!} \lambda \phi^{4}-\frac{i}{4}\lambda \phi^{2}G+\Gamma_{2}[\phi,G;\lambda]\,,\nonumber
\eea

The effect on the theory of changing $\kappa$ is given by the flow equations, which give the derivative of the action (and the $n$-point functions obtained from it) with respect to $\kappa$. Using (\ref{G0-first}) and (\ref{gamma-RG}) gives directly:
\bea
\label{G0RG-a}
\partial_\kappa \hat\Gamma_\kappa = \partial_\kappa \hat\Gamma_{\mathrm{no}\cdot \mathrm{int}\cdot \kappa} = -\frac{1}{2}\partial_ \kappa \hat R_\kappa(G+\phi^2)\,.
\eea

We can obtain this result in a different way by noticing that since the relations between the expectation values and sources are $\kappa$ dependent, the sources will depend on $\kappa$ when the expectation values are taken to be the independent variables. Using (\ref{expt-values-RG}, \ref{legendre-transform-RG}) we obtain
\bea
\label{W-Gamma}
\partial_\kappa \hat\Gamma_\kappa = \bigg[\partial_\kappa W_\kappa +\frac{\delta W_\kappa}{\delta J}\frac{\partial J}{\partial \kappa} +\frac{\delta W_\kappa}{\delta J_2}\frac{\partial J_2}{\partial \kappa}\bigg] - \frac{\partial J}{\partial \kappa} \phi -\frac{\partial J_2}{\partial \kappa}\frac{1}{2}(G+\phi^2) = \partial_\kappa W_\kappa \,,
\eea
and differentiating (\ref{ZandW-RG}, \ref{ZandW-RG-2}) we find the flow equation for the generating functional
\bea
\label{wder}
\partial_\kappa W_\kappa = -\frac{1}{2}\partial_\kappa \hat R_\kappa \langle \varphi \varphi\rangle\,.
\eea
Using the last equality in (\ref{expt-values-RG}) we recover (\ref{G0RG-a}). 

It is useful to define an effective action that corresponds to the original classical action at the scale $\mu$:
\bea
\label{hatGamma}
\Gamma_\kappa = \hat\Gamma_\kappa -\Delta S_\kappa(\phi)\,.
\eea
To make the equations look nicer we define an imaginary regulator function $R_\kappa = -i \hat R_\kappa$ (the extra factor $i$ will be removed when we rotate to Eucledian space to do the numerical calculation). 
Using this notation (and the generic definition $\Phi = i\Gamma$) we have
\bea
&& \Phi_\kappa = \Phi_{\mathrm{no}\cdot \mathrm{int}\cdot\kappa}+ \Phi_{{\rm int}\cdot\kappa}\,,\\
\label{G0-RG}
&& G_{\mathrm{no}\cdot \mathrm{int}\cdot\kappa}^{-1} = G_{\mathrm{no}\cdot \mathrm{int}}^{-1}-R_\kappa \,,\\[2mm]
&& \Phi_{\mathrm{no}\cdot \mathrm{int}\cdot \kappa} = -\big[\frac{1}{2}{\rm Tr}\,{\rm ln} G^{-1}+\frac{1}{2} G_{\mathrm{no}\cdot \mathrm{int}\cdot\kappa}^{-1}G\big]
- \big[\frac{1}{2}G_{\mathrm{no}\cdot \mathrm{int}\cdot\kappa}^{-1} + \frac{1}{2} R_\kappa\big]\phi^2\,,
\eea
and the flow equation (\ref{G0RG-a}) takes the form
\bea
\label{Gflow1}
\partial_\kappa \Phi_\kappa = \frac{1}{2}\partial_\kappa R_\kappa\,\big( \langle \varphi^2\rangle-\phi^2 \big) = \frac{1}{2}\partial_\kappa R_\kappa\,G\,.
\eea
This result has the same form for any $n$PI effective action. The difference in the flow equations for different effective actions is contained in the definition of the expectation values.

\subsection{Flow equations for $n$-point functions}
\label{2PI-hiarchy}

First we derive flow equations for $n$-point functions that would be obtained from the 1PI effective action. Using 
\bea
G = \langle \varphi^2\rangle -\phi^2 = -i\frac{\delta^2 W_{{\rm 1PI}\cdot \kappa}}{\delta J^2} = -\left[\;\frac{\delta^2\hat\Phi_{{\rm 1PI}\cdot\kappa }}{\delta^2\phi}\;\right]^{-1} = -\left[\; \frac{\delta^2 \Phi_{{\rm 1PI}\cdot \kappa }}{\delta\phi^2}+R_k\;\right]^{-1}\,,
\eea
equation (\ref{Gflow1}) becomes
\bea
\label{Gflow2}
\partial_\kappa \Phi_{{\rm 1PI}\cdot\kappa} = -\frac{1}{2} \partial_\kappa R_\kappa \left[\; \frac{\delta^2 \Phi_{{\rm 1PI}\cdot\kappa }}{\delta\phi^2}+R_k\;\right]^{-1}\,.
\eea
Taking functional derivatives of this expression with respect to the expectation value $\phi$ produces the well known infinite hierarchy of functional renormalization group equations. Practical calculations require a truncation of this hierarchy and there is a priori no clear way to decide how to perform this truncation. 

\vspace*{2mm}

Using a similar method we can obtain flow equations for the kernels obtained from the regulated 2PI effective action. We start as before with (\ref{Gflow1}) but now the function $G$ on the right side is an independent functional argument of the effective action which does not depend on $\kappa$. Functionally differentiating (\ref{Gflow1}) produces an infinite hierarchy of coupled equations for the flow of the kernels which will be defined as in (\ref{kernels}). 
As before we consider only the symmetric theory for which $\tilde \phi=0$. The self-consistent 2-point function which solves the equation of motion does depend on $\kappa$, and therefore we do not need to use a tilde to denote the self-consistent solution (as we did in the 2PI calculation), but write it instead as $G_\kappa$. 
We define ($\kappa$ dependent) kernels as in equation (\ref{kernels})
\bea
\label{kernels-RG}
\Phi_{\mathrm{int}\cdot\kappa}^{(n,m)} = 2^m \frac{\delta^n}{\delta \phi^n}\frac{\delta^m}{\delta G^m} \Phi_{\mathrm{int}}\bigg|_{\stackrel{G=G_\kappa}{\phi=o}}\,.\nonumber
\eea
The names of the 2-point and 4-point kernels that we will use below are the same as those defined in (\ref{kernels-specific}).
The stationary condition (see (\ref{stat-cond})) gives the Dyson equation (see (\ref{dyson1})). We define:
\bea
\label{dyson1-RG}
&& G_\kappa^{-1}  =  G_{\mathrm{no}\cdot \mathrm{int} \cdot \kappa}^{-1}-\Sigma_\kappa^{01}(\phi,G_\kappa)\,,\\
\label{dyson1-RG-b}
&& {\bf G}_\kappa^{-1} =G_\kappa^{-1}+R_\kappa =  G_{\mathrm{no}\cdot \mathrm{int}}^{-1} -\Sigma_\kappa^{01}(\phi,G_\kappa)\,.
\eea
A useful relation that we will need later is obtained from differentiating the self-consistent propagator with respect to $\kappa$:
\bea
\label{dif-G}
\partial_\kappa G_\kappa = -G_\kappa \,\partial G_\kappa^{-1}\,G_\kappa  = G_\kappa\,\partial_\kappa(R_\kappa+\Sigma^{01}_{\kappa})\,G_\kappa\,.
\eea

 Using this notation we now derive the 2PI flow equations. Taking functional derivatives of (\ref{Gflow1}) we obtain:
\bea
\label{RGhi-1}
2^m \bigg[\partial_\kappa\frac{\delta^n}{\delta \phi^n}\frac{\delta^m}{\delta G^m} \Phi_\kappa \bigg]_{\stackrel{G=G_\kappa}{\phi=o}}
= \frac{1}{2}2^m \bigg[\frac{\delta^n}{\delta \phi^n}\frac{\delta^m}{\delta G^m} \bigg(\int dQ\, \partial_\kappa R_\kappa(Q) \, G\bigg)\bigg]_{\stackrel{G=G_\kappa}{\phi=o}}\,.
\eea
The left side of the equation can be written $\big(\partial_\kappa[f(\kappa,G)]\big)_{\stackrel{G=G(\phi)}{\phi=o}}$. In order get something that can be written as a kernel, we need to obtain an expression in which the derivative with respect to $\kappa$ is taken after the self-consistent solutions are substituted. This can be done using the chain rule since for any function $f(\kappa,G)$
\bea
\label{any-fcn}
\partial_\kappa f(\kappa,G_\kappa)  = \bigg[\partial_\kappa f(\kappa,G)\bigg]_{\stackrel{G=G_\kappa}{\phi=o}} +\partial_\kappa G_\kappa \;\frac{\delta f(\kappa,G)}{\delta G}\bigg|_{\stackrel{G=G_\kappa}{\phi=o}}\,.
\eea
Using (\ref{RGhi-1}, \ref{any-fcn}) we obtain
\bea
\label{RGhi-2}
\!\!\!\!\!\!\!\!\!\!\!\!\!\!\!\!&& \frac{1}{2}2^m \bigg[\frac{\delta^n}{\delta \phi^n}\frac{\delta^m}{\delta G^m} \bigg(\int dQ\, \partial_\kappa R_\kappa(Q) \, G\bigg)\bigg]_{\stackrel{G=G_\kappa}{\phi=o}}
=\partial_\kappa \Phi_{{\rm int}\cdot\kappa}^{(n,m)}-\frac{1}{2}\partial_\kappa G_\kappa  \Phi_{{\rm int}\kappa}^{(n,m+1)} +e_1+e_2\,, \\[4mm]
\!\!\!\!\!\!\!\!\!\!\!\!\!\!\!\!&& e_1= 2^m \partial_\kappa  h^{(n,m)}\,,~~~~e_2 = -\partial_\kappa G_\kappa 2^m h^{(n,m+1)}\,,~~~~ h^{(n,m)} = 2^m \frac{\delta^n}{\delta \phi^n}\frac{\delta^m}{\delta G^m} \Phi_{\mathrm{no}\cdot\mathrm{int}\cdot\kappa}\bigg|_{\stackrel{G=G_\kappa}{\phi=o}}.\nonumber
\eea
It is easy to show that the contribution of the term $-\frac{1}{2}{\rm Tr}\,{\rm ln} G^{-1}$ in $\Phi_{\mathrm{no}\cdot\mathrm{int}\cdot\kappa}$ to the sum $e_1+e_2$ is zero. 
For $(n,m)\ne (0,1)$, the left side of (\ref{RGhi-2}) is zero, and 
$e^\prime_1=e^\prime_2=0$ (where the primes indicate that the log term has been dropped from $\Phi_{\mathrm{no}\cdot\mathrm{int}\cdot\kappa}$). 
For $(n,m) =  (0,1)$ we have (left side) = $e^\prime_1$ and $e^\prime_2=0$. 
The result is that for any values of $(n,m)$ we can drop the left side of (\ref{RGhi-2}) and the two terms $e_1$ and $e_2$. 
Using (\ref{dif-G}) equation (\ref{RGhi-2}) can be written:
\bea
\label{RGhi-3}
\partial_\kappa \Phi_{{\rm int}\kappa}^{(n,m)}\bigg|_{\stackrel{G=G_\kappa}{\phi=o}} 
= \frac{1}{2}\partial_\kappa \,(R_\kappa+\Sigma^{01}_\kappa)
\,G_\kappa^2\, \partial_\kappa \Phi_{{\rm int}\kappa}^{(n,m+1)}\bigg|_{\stackrel{G=G_\kappa}{\phi=o}}\,.
\eea
Equation (\ref{RGhi-3}) gives a series of infinite hierarchies of coupled equations for the 2PI kernels in which kernels with fixed $n$ and different $m$ are coupled together. 
However, unlike the hierarchy produced from the 1PI effective action, when the 2PI effective action is truncated at some finite loop order, the hierarchy in (\ref{RGhi-3}) is also truncated. 

\vspace*{2mm}

We note that the flow equations for the 2-point kernels are special in the sense that they can be rewritten in terms of BS vertices. 
For $(n,m)=(0,1)$ equation (\ref{RGhi-3}) gives (using (\ref{kernels-specific}, \ref{dyson1-RG})) 
\bea
\label{solve-M}
\partial_\kappa G_\kappa^{-1} && = -\partial_\kappa R_\kappa + \frac{1}{2} \partial_\kappa G_\kappa^{-1} G_\kappa^2  \Lambda_\kappa^{02} \,.
\eea
This expression can be rewritten in terms of the BS vertex $M$ by expanding it to obtain
\bea
\partial_\kappa G_\kappa^{-1}&& = -\partial_\kappa R_\kappa-\frac{1}{2}\partial_\kappa R_\kappa G_\kappa^2
\big(\Lambda_\kappa^{02} + \frac{1}{2}\Lambda_\kappa^{02} G_\kappa^2 \Lambda_\kappa^{02} +\frac{1}{4}\Lambda_\kappa^{02} G_\kappa^2 \Lambda_\kappa^{02}G_\kappa^2 \Lambda_\kappa^{02}+\cdots\big) \,, \nonumber\\
 &&= -\partial_\kappa R_\kappa-\frac{1}{2}  \partial_\kappa R_\kappa \,G_\kappa^2 \, M_\kappa\,,\nonumber\\[2mm]
\label{flow-sigma01}
\to~~~~~ \partial_\kappa \Sigma_\kappa^{01} && = \frac{1}{2} \partial_\kappa R_k\,G^2_\kappa \,M_\kappa\,,
\eea
where we have used (\ref{M-defn}) in the second line and (\ref{dyson1-RG}) in the third. 

For $(n,m)=(2,0)$ equation (\ref{RGhi-3}) gives
\bea
\partial_\kappa \Sigma_\kappa^{20} = -\frac{1}{2}\partial_\kappa G_\kappa^{-1}\,G_\kappa^2\; \Lambda_\kappa^{21}\,.
\eea
Substituting (\ref{solve-M}) on the right side, introducing the vertex $M^\prime$ defined in (\ref{Mprime2-defn}), and using the fact that the vertices $\Lambda^{21}_\kappa$, $M$ and $M^\prime$ are symmetric, we obtain
\bea
\label{flow-sigma20}
\partial_\kappa \Sigma_\kappa^{20} && = \frac{1}{2}(\partial_\kappa R_\kappa+\frac{1}{2}\partial_\kappa R_\kappa G_\kappa^2 M_\kappa)G_\kappa^2 \Lambda_\kappa^{21}
 = \frac{1}{2}\Lambda_\kappa^{21} G_\kappa^2 \partial_\kappa R_\kappa + \frac{1}{4}\Lambda_\kappa^{21}G_\kappa^2 M_\kappa G_\kappa^2 \partial_\kappa R_\kappa \\[2mm]
&&  = \frac{1}{2}(\Lambda_\kappa^{21}+\frac{1}{2}\Lambda_\kappa^{21} G_\kappa^2 M_\kappa)G_\kappa^2\partial_\kappa R_\kappa
 = \frac{1}{2} M_\kappa^\prime G_\kappa \partial_\kappa R_\kappa
 = \frac{1}{2}  \partial_\kappa R_\kappa  G_\kappa^2 \, M_\kappa^\prime \,.\nonumber
\eea

\subsection{FRG Flow Equations at order $\lambda^2$}
\label{flow-specific}

In this section we derive the flow equations obtained from 
the effective action by including terms up to order $\lambda^2$ in the skeleton expansion. We consider only $\Sigma^{01}_\kappa$, $\Lambda^{02}_\kappa$ and $M_\kappa$, and we drop the superscripts on the self-energy and the 4-kernel. 

We rotate to Eucledian space using Eq. (\ref{euc-defn}) and define $R_\kappa=-i R_{\kappa\,E}$ and $\Lambda^{03}_\kappa = -i \Lambda^{03}_{\kappa E}$ (the extra factors of $i$ remove the factors that were introduced in the definitions $\lambda_{\rm phys}=i\lambda$ (under equation (\ref{action})) and $\hat R=i R$ (under equation (\ref{hatGamma}))). 
From this point forward we suppress the subscripts which indicate Eucledian space quantities. In Table \ref{2-point-fcn-table} we summarize the definitions of the different Eucledian 2-point functions used in this paper.

\begin{table}
\begin{center}
\begin{tabular}{|c|l |} 
\hline
$G_{\mathrm{no}\cdot \mathrm{int}}$ & the non-interacting propagator  \\\hline

$G_{\mathrm{no}\cdot \mathrm{int}\cdot\kappa}$ & defined as $G^{-1}_{\mathrm{no}\cdot \mathrm{int}\cdot\kappa} = G^{-1}_{\mathrm{no}\cdot \mathrm{int}} + R_\kappa$ \\\hline

$G$ & the functional argument of the effective action\\\hline



$G_\kappa$ & the self-consistent solution (called $\tilde G$ in the 2PI section)~~ \\\hline

${\bf G}_\kappa$ & defined by ${\bf G}_\kappa^{-1} = G_\kappa^{-1} - R_\kappa$\\\hline

$G_\mu$ & the boundary value at $\kappa=\mu$ from which the flow starts \\\hline

$G_0$ & the solution of the flow equations at $\kappa=0$ \\\hline

$\Sigma_\kappa$ & defined by $G_\kappa^{-1} = G_{\mathrm{no}\cdot \mathrm{int}\cdot\kappa}^{-1} + \Sigma_\kappa$  or ${\bf G}_\kappa^{-1} = G_{\mathrm{no}\cdot \mathrm{int}\cdot}^{-1} + \Sigma_\kappa$\\\hline

\end{tabular}
\end{center}
\caption{Definitions of different Eucledian 2-point functions\label{2-point-fcn-table}}
\end{table}

In Eucledian space (\ref{solve-M}) can be rewritten (using (\ref{dyson1-RG-b})) as
\bea
\label{flow-sigma2}
\partial_\kappa\Sigma_\kappa(P) = \frac{1}{2}\int dQ \partial_\kappa\big(\Sigma_\kappa(Q) + R_\kappa(Q)\big)G_\kappa^2(Q)\Lambda_\kappa(Q,P)\,.
\eea
The flow equation for the 4-point kernel (equation (\ref{RGhi-3}) with $(n,m)=(0,2)$) is 
\bea
\label{flow-lambda2}
\partial_\kappa \Lambda_\kappa(P,K) = \frac{1}{2}\int dQ \,\partial_\kappa \big[R_\kappa (Q)+\Sigma_\kappa (Q)\big] G^2_\kappa(Q) \Lambda^{03}_\kappa(Q,P,K)\,.
\eea
Using $(n,n)=(0,3)$ can write an equation for the flow of the kernel $\Lambda_\kappa^{03}$, but it will not be needed. This is explained in section \ref{truncation-section}.

\subsection{Boundary Conditions}
\label{boundary-conditions-section}
In order to solve these flow equations, one must specify the boundary conditions from which the flow starts at $\kappa=\mu$. These boundary conditions must be consistent with the renormalization conditions that we want to impose at the $\kappa=0$ end of the flow. 
In order to compare with the results obtained from the 2PI calculation we use the same renormalization conditions which now take the form (see (\ref{renorm-conds2})):
\bea
\label{renorm-conds3}
&& {\bf G}_0^{-1}(0) =m^2\,,~~ \frac{d}{dP^2}{\bf G}_0^{-1}\bigg|_{P=0} =1\,,~~  M_0(0) =- \lambda \,.
\eea
We remind the reader that the subscripts 0 indicate $\kappa=0$, and not bare or non-interacting quantities. 
In addition, we need to show that all divergences can be absorbed into the definitions of the fundamental parameters at the scale $\kappa=\mu$. 

We start with the flow equation for the 2-point function. The solution of the differential equation  (\ref{flow-sigma2}) gives the $\kappa$ dependent 2-point function, up to an integration constant:
\bea
\partial_\kappa \Sigma_\kappa(P) ~~\to~~ \Sigma_\kappa(P) + C(P)\,,
\eea
where $C(P)$ is any function that does not depend on $\kappa$. 
Using the Dyson equation in Eucledian space and choosing $C=-(\Sigma_{0}(0)+P^2\Sigma^\prime_{0}(0)\big)$ we have
\bea
\label{Geq1}
{\bf G}^{-1}_\kappa = P^2+m^2+\Sigma_\kappa(P)-\big(\Sigma_{0}(0)+P^2\Sigma^\prime_{0}(0)\big)\,,
\eea
there the prime indicates a derivative with respect to $P^2$. The renormalization conditions are now automatically satisfied at $\kappa=0$, but we have to check that ${\bf G}^{-1}_\mu$ approaches the classical solution when $\mu\to\infty$. We rewrite (\ref{Geq1}) as
\bea
\label{Geq2}
{\bf G}^{-1}_\kappa && =P^2+\big(\Sigma_\kappa(P)-(\Sigma_\kappa(0)+P^2\Sigma^\prime_\kappa(0))\big) \\[2mm]
&& + \big[m^2+\Sigma_\kappa(0)-\Sigma_{0}(0)\big] +P^2 \big[\Sigma_\kappa^\prime(0)-\Sigma_{0}^\prime(0)\big]\,.\nonumber
\eea
The terms in the first square bracket are proportional to a running $\kappa$ dependent mass and the second square bracket corresponds to a running wave-function renormalization constant. We can use these parameters to rewrite the original equation:
\bea
\label{running-mass}
&& m^2+\Sigma_\kappa(0)-\Sigma_{0}(0) = Z_\kappa m_\kappa^2 \,, \\
\to ~ ~ && Z_\mu m_\mu^2 = m^2+\Sigma_\mu(0)-\Sigma_{0}(0)
~~ \Rightarrow ~ ~  \Sigma_{0}(0) = m^2 - Z_\mu m_\mu^2 + \Sigma_\mu(0)\,,\nonumber\\
&&  \Sigma_\kappa^\prime(0)-\Sigma_0^\prime(0) = \delta Z_\kappa \,,\nonumber \\
~~ \to ~ ~ && \delta Z_\mu = \Sigma_{\mu}^\prime(0)-\Sigma_{0}^\prime(0)
\Rightarrow ~ ~  \Sigma_{0}^\prime(0) = -\delta Z_\mu + \Sigma_{\mu}^\prime(0)\,. \nonumber
\eea
Substituting these expressions for $\Sigma_{0}(0)$ and $\Sigma^\prime_{0}(0)$ into (\ref{Geq1}) we obtain
\bea
\label{ass-prop}
{\bf G}_\mu^{-1} = P^2(1+\delta Z_\mu) +Z_\mu  m_\mu^2  + \big[\Sigma_\kappa(P)-\big(\Sigma_{\mu}(0)+P^2 \Sigma^\prime_{\mu}(0)\big)\big]\,.
\eea
In Appendix \ref{appendixB} we show  that the quantity in square brackets approaches zero when $\kappa \to\mu \gg P$ which means that we can use ($Z_\mu = 1+\delta Z_\mu$)
\bea
\label{bc-prop}
{\bf G}_\mu^{-1} = Z_\mu(P^2 + m_\mu^2)
\eea
as the boundary condition on the flow equations. 

Now we consider the boundary condition on the flow equation for the 4-kernel. The solution of the differential equation  (\ref{flow-lambda2}) gives:
\bea
\label{lam1}
\partial_\kappa \Lambda_\kappa(P,Q)  ~~\to~~ \Lambda_\kappa(P,Q) = \Lambda^{\rm loop}_\kappa(P,Q) + C\,.
\eea
In principle, $C$ could contain a function of momentum that does not depend on $\kappa$, but we choose to absorb any momentum dependent contributions into $\Lambda^{\rm loop}_\kappa(P,Q)$. 
We choose $C = -(\lambda+\Delta\lambda+\Lambda^{\rm loop}_0(0,0))$ so that (\ref{lam1}) becomes
\bea
\label{lam1b}
\Lambda_\kappa(P,Q) = -\big(\lambda+\Delta\lambda+\Lambda^{\rm loop}_0(0,0)\big)+\Lambda^{\rm loop}_\kappa(P,Q)\,.
\eea
and $\Lambda_0(0,0) = -(\lambda+\Delta\lambda)$. 
The parameter $\Delta\lambda$ is a constant (independent of $\kappa$ and momentum) which accounts for the fact that we have no reason to require that $\Lambda_0(0,0)$ equals $M_0(0,0)$. 
We use the same trick as before to extract the behaviour of the kernel when $\kappa\to\mu$. We rewrite (\ref{lam1b}) as
\bea
\label{lam2}
\Lambda_\kappa(P,Q) = \big(\Lambda^{\rm loop}_\kappa(P,Q)-\Lambda^{\rm loop}_\kappa(0,0)\big) - \Delta\lambda - \big[\lambda+ \Lambda^{\rm loop}_0(0,0)-\Lambda^{\rm loop}_\kappa(0,0)\big]\,.
\eea
The quantity in square brackets is a running coupling. We define:
\bea
\label{lam2b}
\lambda_\kappa  = \lambda+ \Lambda^{\rm loop}_0(0,0)-\Lambda^{\rm loop}_\kappa(0,0) \,,
\eea
and rewrite the solution of the vertex flow equation at the scale $\mu$ using
\bea
\label{lam3}
&& \lambda_\mu = \lambda + \Lambda^{\rm loop}_0(0,0)-\Lambda^{\rm loop}_\mu(0,0)~~\Rightarrow ~~ 
\Lambda^{\rm loop}_0(0,0) = \lambda_\mu - \lambda  + \Lambda^{\rm loop}_\mu(0,0)\,.\nonumber
\eea
Substituting this result into (\ref{lam1b}) we obtain
\bea
\label{lam4}
\Lambda_\kappa(P,Q) = -\lambda_\mu -\Delta\lambda +\big[\Lambda^{\rm loop}_\kappa(P,Q)-\Lambda^{\rm loop}_\mu(0,0)\big]\,.
\eea
In Appendix \ref{appendixB}
we show that the quantity in square brackets goes to zero for $\kappa \to\mu \gg \{P,Q\}$, except for a momentum independent contribution which can be absorbed into the definition of $\lambda_\mu$. 
Likewise, since $\lambda_\mu$ is a constant that will be tuned to produce the renormalized coupling, we can simply absorb the extra piece $\Delta\lambda$ into its definition. 
The result is that we can use the boundary condition 
\bea
\label{bc-lam}
\Lambda_\mu(P,Q)=-\lambda_\mu\,
\eea in the flow equation (\ref{flow-lambda2}). 

\subsection{Truncation}
\label{truncation-section}

Now we consider solving the flow equations using these boundary conditions. 
Equation (\ref{flow-lambda2}) for $\Lambda_\kappa=\Lambda_\kappa^{02}$ depends on the higher order kernel $\Lambda_\kappa^{03}$. One could write an equation for the flow of the kernel $\Lambda_\kappa^{03}$ of the form $\partial_\kappa \Lambda_\kappa^{03}\sim \int dQ\, \partial_\kappa G_\kappa \,\Lambda_\kappa^{04}$. At the level of our approximation however, the kernel $\Lambda_\kappa^{04}$ is a constant, and therefore the right side of the equation for $\partial_\kappa \Lambda_\kappa^{03}$ is an exact differential which can be integrated directly. The integration constant must be set to zero because there is no 6-vertex in the Lagrangian. Equivalently, one can simply obtain $\Lambda_\kappa^{03}$ directly from the effective action using  (\ref{kernels-RG}):
\bea
\label{LAM03-RG}
&&\Lambda^{03}(Q,P,K) \\
~~~~~&&= -\lambda^2\big(G_\kappa(Q+P+K)+G_\kappa(Q+P-K)+G_\kappa(Q-P+K)+G_\kappa(Q-P-K)\big)\,.\nonumber
\eea

The equations (\ref{flow-sigma2}, \ref{flow-lambda2}, \ref{LAM03-RG}) can be solved simultaneously using the boundary conditions (\ref{bc-prop}, \ref{bc-lam}). 
The momentum integral in (\ref{flow-lambda2}) is completely finite because of the structure of the kernel $\Lambda_\kappa^{03}$ in (\ref{LAM03-RG}). 
The momentum integral in (\ref{flow-sigma2}) is finite except for a momentum independent piece produced by a constant term in the 4-kernel. This divergence can be absorbed into the definition of $m_\mu$ (see the discussion on the definition of the coupling $\lambda_\mu$ above (\ref{bc-lam})). 
The result is therefore that all of the divergent contributions have been absorbed into the definitions of the parameters $m_\mu$ and $\lambda_\mu$.
For a given choice of the function $R_\kappa$, the theory is completely specified by the flow equations and the initial conditions, and one may ``forget'' their origins from a functional integral.

In fact, since the integral on the right side of the flow equation (\ref{flow-lambda2}) is finite and the integration constant is known, we do not need to solve the flow equation for $\Lambda_\kappa$, but can write directly:
\bea
\label{wj-lambda1}
\Lambda_\kappa(P,Q) =-\lambda_\mu+\frac{\lambda^2}{2}\int dQ\, G_\kappa(Q)\big[G_\kappa(Q+P-K) + G_\kappa(Q+P+K)\big]\,.
\eea
It is easy to see that this expression satisfies (\ref{flow-lambda2}) together with (\ref{LAM03-RG}). One can also verify that (\ref{wj-lambda1}) satisfies the boundary condition (\ref{bc-lam}), by showing that in the limit $\mu\gg \{P,K\}$ the integral reduces to a constant, which can be absorbed into the definition of $\lambda_\mu$. This can be done following the method in Appendix \ref{appendixB}. 

We note that (\ref{wj-lambda1}) is just the 2PI kernel with the tree term replaced by the vertex $\lambda_\mu$ (see equation (\ref{hatLam-discrete})). 
However, we can {\it not} start from some kind of similarly modified 2PI expression for $\Sigma_\kappa$, because the flow equation for the 2-point function (\ref{flow-sigma2}) contains a embedded sub-divergence which cannot be removed unless $\Lambda_\kappa$ is calculated self-consistently from its flow equation. 

\subsection{Connection to the 2PI formalism.}
\label{2pi-connection}

In this section we discuss the connection between the RG formalism and the standard 2PI calculation. 
From equations (\ref{running-mass}, \ref{lam2b}) we see that in the limit $\mu\to\infty$ the running mass and running coupling can be written
\bea
&& \lim_{\mu\to\infty} Z_\mu m_\mu^2 \to Z m_b^2  =   m^2 +\delta m^2~~{\rm with}~~ \delta m^2 = -\big(\Sigma_0(0)-\Sigma_{\mu\to\infty}(0)\big)\,,\\
&& \lim_{\mu\to\infty}\lambda_\mu \to \lambda_b = \lambda +  \delta\lambda \,,~~{\rm with}~~ \delta \lambda = \big(\Lambda^{\rm loop}_0(0,0)-\Lambda^{\rm loop}_{\mu\to\infty}(0,0)\big)\,.
\eea
Comparing with the counter-terms used in the 2PI calculation (\ref{32}, \ref{35}) (using $\Sigma(0)=\delta m^2+\hat\Sigma(0)$ and $\Lambda^{\rm loop} = \lambda+\hat\Lambda$ from (\ref{sigma-discrete}, \ref{hatLam-discrete})), these expressions contain extra momentum independent contributions ($\hat\Sigma_{\mu\to\infty}(0)$ and $\Lambda^{\rm loop}_{\mu\to\infty}(0,0)$) which are simply absorbed into the definitions of the bare parameters by the tuning process. 
The conclusion is that if we impose the renormalization conditions on the 2-point function and BS vertex at the $\kappa=0$ end of the flow, then in the limit $\mu\to\infty$ the constants of integration which appear in the solutions of the flow equations play the role of the counter-terms that are introduced in the 2PI formalism.

It is interesting to consider what would happen to this structure at next order in the approximation. 
In section \ref{renorm-2PI-section} we discussed the fact that in the 2PI formalism at arbitrary order, the coupling counter-term must be divided into two pieces: $\delta\lambda = \delta\lambda^\prime+\Delta\lambda$, and these two pieces must be determined from two different renormalization conditions, on $\Lambda$ and $M$.
When we truncate at the level of the basket-ball diagram, as in this paper, the divergence in $\Lambda$ is global, and  the two counter-terms can be combined in the numerical calculation and one needs only to find $\delta\lambda$. 
In the RG calculation, truncating at the level of the basket-ball diagram means that we do not have to solve the flow equation for the vertex $\Lambda_\kappa$ but can integrate it directly to obtain (\ref{wj-lambda1}). 
At next order in the truncation of the effective action, one would need two renormalization conditions in the 2PI calculation and two flow equations in the RG calculation. In this regard, neither calculation is easier than the other.

Finally, we comment on the fact that the 4-vertex that must satisfy the renormalization condition of the form $V(0)=-\lambda$ is the vertex $M$, and not, for example, either the 4-kernel $\Lambda$ or the resummed vertex $\tilde\Phi^{(4)}$. This point was understood in the 2PI calculation by diagrammatic analysis of the overlapping sub-divergences that appear when the integral equations for the 2-point and 4-point functions are iterated. In contrast, in the RG approach, the renormalization condition on the 4-vertex $M$ emerges naturally from the flow equations (see equation (\ref{flow-sigma01})). The structure of the sub-divergences in higher order effective actions (3PI, 4PI, etc) is too difficult to untangle using diagrammatic analysis. The RG approach we have outlined in this paper provides a promising alternative method to study the renormalizability of these theories.

\vspace*{2mm}

\subsection{Numerical Method}
\label{RG-numerical-section}

The flow equations can be discretized as in section \ref{discretization-section}. The discretized forms of equations (\ref{flow-sigma2}) and (\ref{wj-lambda1}) are
\bea
\label{wj-lam-eqn}
&&\Lambda({\bf n},{\bf l}) = -\lambda_\mu + \frac{\lambda^{2}}{2} \,\sum_{\bf j}
G({\bf j})G({\bf j} + {\bf n} - {\bf l})  +\frac{\lambda^{2}}{2} \,\sum_{\bf j}
G({\bf j})G({\bf j} + {\bf n} + {\bf l})\,, \\
\label{sigma-discrete-RG}
&&\partial_\kappa\Sigma_\kappa({\bf n}) = \frac{1}{2}\sum_{\bf j} \partial_\kappa\big(\Sigma_\kappa({\bf j}) + R_\kappa({\bf j})\big)G^2({\bf j})\Lambda_\kappa({\bf j},{\bf n})\,.
\end{eqnarray}

We do not tune the parameter $Z_\mu$ but instead follow \cite{Berges2002,BMW2} and use a $\kappa$ dependent wave function renormalization which guarantees that no intrinsic scale is introduced in the average inverse propagator. 
We use \cite{Berges2002,BMW2}
\bea
&& R_\kappa({\bf n}) = Z_\kappa \frac{Q^2}{e^{\frac{Q^2}{\kappa^2}}-1}\,,~~Q^2=m_t^2 n_4^2+m_s^2(n_1^2+n_2^2+n_3^2)\,,\\
\eea
where $Z_{\kappa}$ is defined below.

To compare with the 2PI calculation we use the same parameters as in the previous section. We choose the renormalized parameters as $m=1$ and $\lambda=1$ and renormalize at $N_t=128$, $N_s=32$, $a_t=1/16$ and $L=a_s N_s=2$. 
We use $\mu=100$ and solve the flow equations in $N_\kappa=50$ steps down to $\kappa_{\rm min}=0.01$. 
After tuning, the final values of the tuned bare parameters are $m_\mu^2=-10.7541$ and $\lambda_\mu = 1.1076$. These values are not initially known, and the tuning loop converges fastest if the initial guess is close to the final solution. 


To solve the flow equations we follow the steps below:

\begin{enumerate}

\item Start at the scale $\kappa=\mu$ from the values determined by the boundary conditions: ${\bf G}_\mu^{-1}({\bf n}) = \big(m^2_t n_4^2 + m_s^2(n_1^2+n_2^2+n_3^2)+m_\mu^2\big)$ and $\Lambda_\mu = -\lambda_\mu$. 

\item Relax (\ref{sigma-discrete-RG}) to find $\Delta\Sigma_\mu = \delta\kappa \cdot (\partial_\kappa\Sigma_\kappa)|_{\kappa=\mu}$. The relaxation procedure is exactly the same as in the 2PI calculation (see section \ref{2PI-iterations-section}). We start with an input function ($\Delta\Sigma^{(0)}_\mu=0$) on the right side of (\ref{sigma-discrete-RG}) and compute the output function as the left side. The output function is then used as the input for the next iteration. 
We continue until the output function agrees with the input function, to within $10^{-4}$.

\item Find ${\bf G}^{-1}_{\mu-\delta\kappa} = {\bf G}^{-1}_\mu + \Delta\Sigma_\mu$.

\item Find $Z_{\mu-\delta\kappa} =  \big[{\bf G}^{-1}_{\mu-\delta\kappa}(0,0,0,1)-{\bf G}^{-1}_{\mu-\delta\kappa}({\bf 0})\big]\frac{1}{m_s^2}$\,.

\item Find $G_{\mu-\delta\kappa}=Z_{\mu-\delta\kappa}/\big[{\bf G}^{-1}_{\mu-\delta\kappa}+Z_{\mu-\delta\kappa}R_{\mu-\delta\kappa} \big]$\,.

\item Set $\Delta\Sigma^{(0)}_{\mu-\delta\kappa} = \Delta\Sigma_\mu$ as the initial value to start the relaxation for the next cycle.

\item Use (\ref{wj-lam-eqn}) to find $\Lambda_{\mu-\delta\kappa}$. 

\item $\cdots$ using $G_{\mu-\delta \kappa}$ and $\Lambda_{\mu-\delta \kappa}$ and $\Delta\Sigma^{(0)}_{\mu-\delta\kappa}$ repeat  steps (2 - 7) $N_\kappa$ times and arrive at $G_0$ and $\Lambda_0$.  

\item 
Using $G_{0}$ and $\Lambda_{0}$ relax (\ref{M-discrete}) to obtain $M_0$. 

\item Find the output (renormalized) parameters
\bea
\label{renorm-conds5}
&& m^2_{\rm found} = G_0^{-1}(0)\,,\\
&&\lambda_{\rm found} = -M_{0}(0,0)\,.
\eea
If the ``found'' renormalized parameters match the chosen ones ($m^2_{\rm goal}=\lambda_{\rm goal}=1$ in our calculation), the  calculation is finished. Save the corresponding values of $m_\mu$ and $\lambda_\mu$ to use in the finite temperature calculation. If the renormalized parameters do not equal the chosen goals (to within $10^{-4}$) continue to the next step.

\item Update the bare parameters using ($\alpha=.6$)
\bea
&& m^{2\prime}_{\mu} = m^2_\mu+\alpha(m^2_{\rm goal}-m^2_{\rm found})\,,\\
&& \lambda^{\prime}_{\mu} = \lambda_\mu+\alpha(\lambda_{\rm goal}-\lambda_{\rm found})\,.
\eea

\item Remove the primes from the bare parameters and repeat the whole calculation starting from step (1). 
\end{enumerate}

At finite temperature we repeat steps (1 - 10) using the bare parameters $m_\mu$ and $\lambda_\mu$ that were obtained from the zero temperature calculation.

In practice it is convenient to solve the differential equations using a logarithmic scale 
\bea
&& t=\ln \kappa/\mu\,,~~\kappa\partial_\kappa = \partial_t \,,\\
&& t = n_\kappa \,\delta t\,,~~~
\delta t = \ln[\kappa_{\rm min}/\mu]/N_{\kappa}\,.
\eea

\newpage

\section{Numerical Results}
\label{results-all}

We use $a_t=1/16$, $a_s=1/16$, $N_s=32$ and renormalize the theory at $T_0= (N_t^{\rm max} a_t)^{-1}$.  The maximum value of $N_t$ is limited by computation time and memory. We use $N_t^{\rm max}$=128 which gives $T_0=0.125$. 
In the RG calculation we use $\mu=100$ and solve the flow equations in $N_\kappa=50$ steps down to $\kappa_{\rm min}=0.01$. 
We have checked that the RG results are unchanged if $\kappa_{\rm min}$ is reduced or $\mu$ is increased. 
To study finite temperatures we use a range of values for $N_t$ such that
\bea
\label{mass-scales}
T_0 =\big[a_t N_t^{\rm max}\big]^{-1} \ll m < T_{\rm max}=\big[a_t N_t^{\rm min}\big]^{-1}\,.
\eea
We use $N_t^{\rm min}=8$ which gives $T_{\rm max}=2m$. 

Figure \ref{GMfig} shows the inverse 2-point function and Bethe-Salpeter vertex at zero momentum as functions of temperature. The 2PI and RG calculations agree well, which shows that the 2PI calculation can be done without counter-terms by using a RG regulator and solving the flow equations. 
From the plot of $G^{-1}(0)$ versus $T$ we see that 
\bea
T_0\ll T^*\equiv m(T^*) \approx 1\,,
\eea
which verifies that $N_t=128$ can be taken as the zero temperature limit. 
\begin{figure}
\centering
\mbox{\subfigure{\includegraphics[width=3.5in]{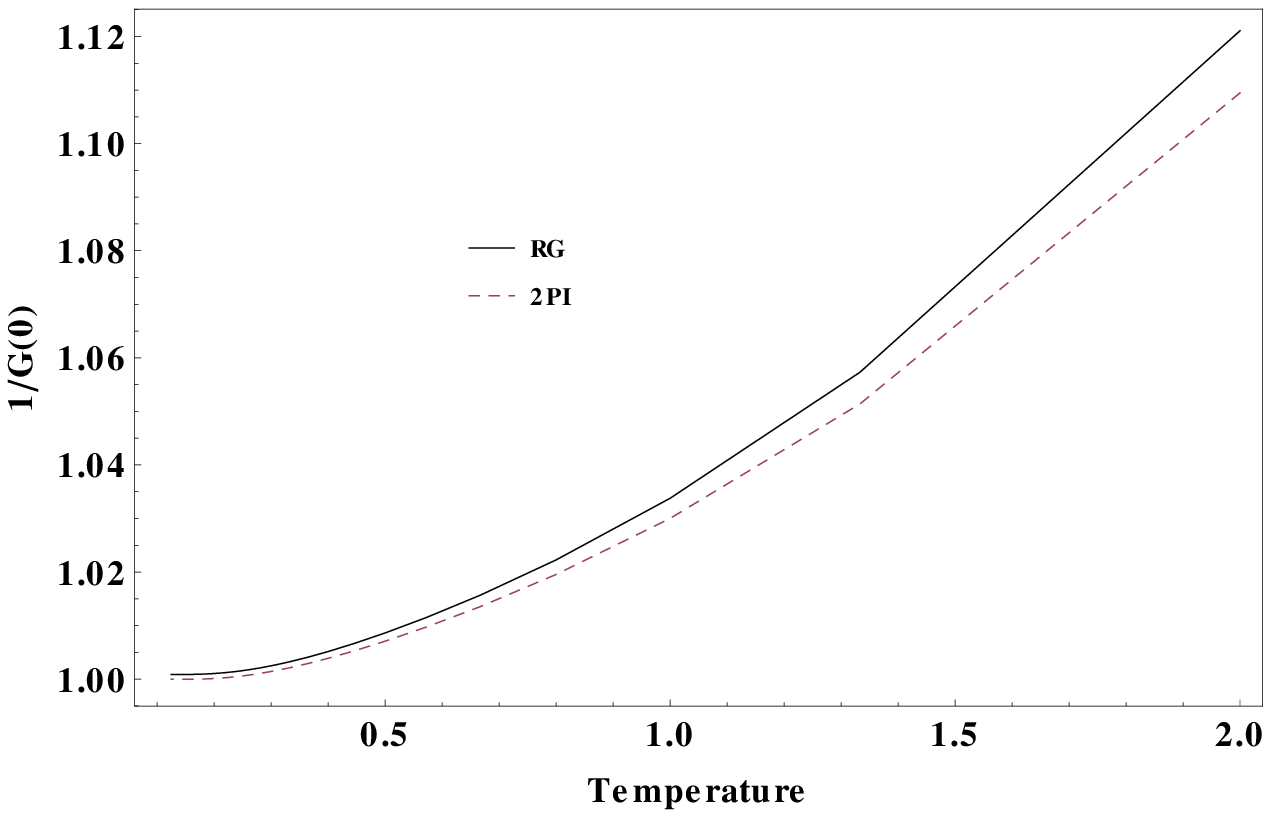}}\quad
\subfigure{\includegraphics[width=3.5in]{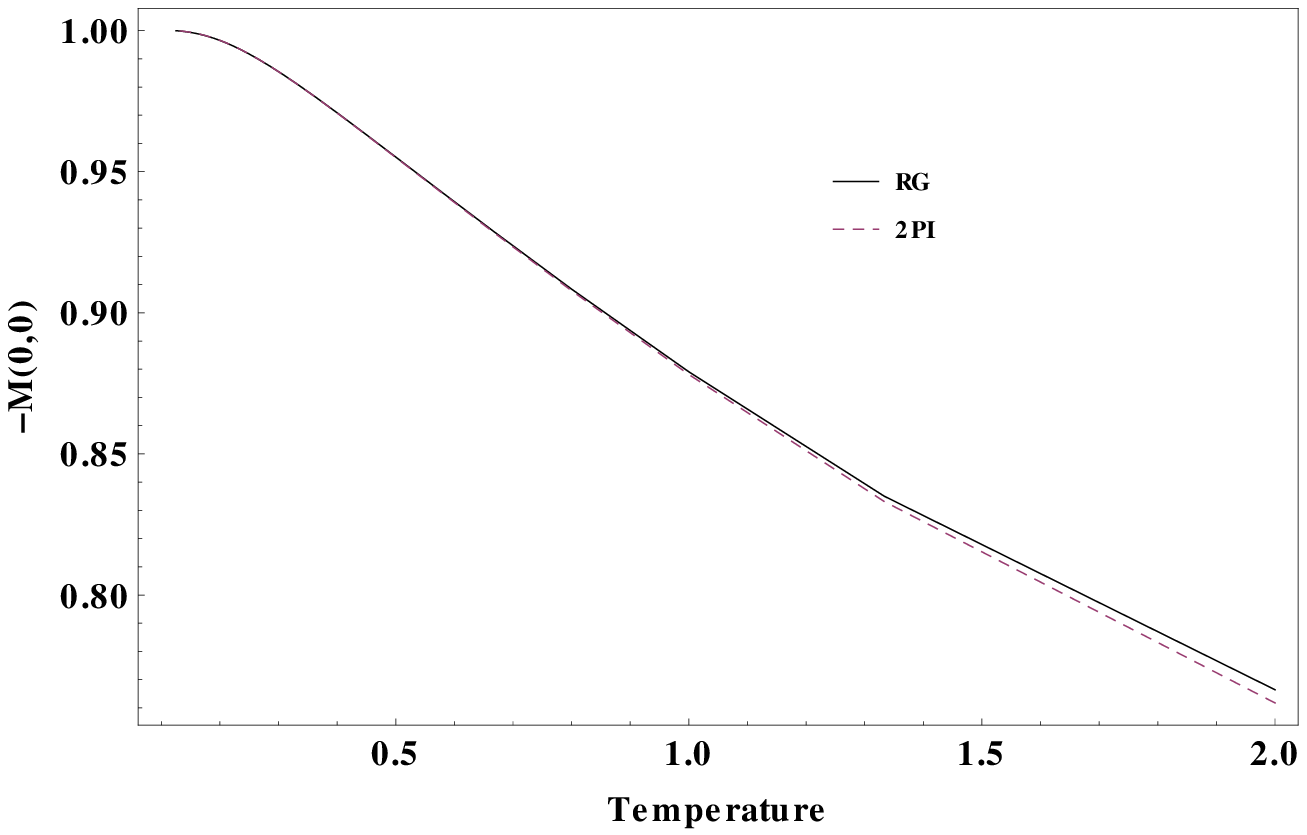} }}
\caption{The inverse propagator and Bethe-Salpeter vertex at zero momentum as functions of the temperature, from the 2PI and RG calculations. 
\label{GMfig}}
\end{figure}

We test the renormalization by reducing the lattice spacing in the spatial direction ($a_s$) while holding the spatial length of the box ($L=a_s N_s$) fixed. In Fig. \ref{as-plot} we plot $M(0,0)$ versus $\log 1/a_s$. For comparison, we repeat the 2PI calculation using an incorrect renormalization procedure, by adding vertex counter-terms ($\lambda \to \lambda+\delta\lambda$) to the basketball diagram (see Fig. \ref{phi-fig}).  The graph shows that in the incorrect calculation, $M(0,0)$ increases when $a_s$ is reduced. The 2PI calculation is almost flat, and the RG calculation is flatter still, which shows that the renormalization is done correctly. 
\begin{figure}
\begin{center}
\includegraphics[width=9cm]{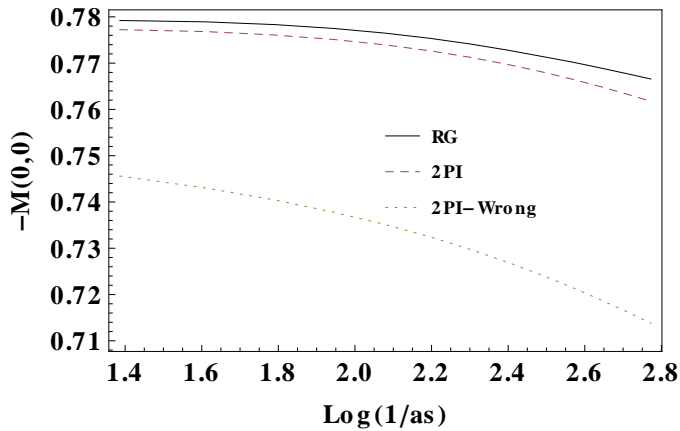}
\end{center}
\caption{The dependence of $M(0,0,0,0)$ on the lattice separation in the spatial directions, for the 2PI calculation, the RG calculation, and an incorrect version of the 2PI calculation which is included for comparison. \label{as-plot}}
\end{figure}

\section{Conclusions}
\label{conclusions-section}

The 2PI-RG calculation is slower than the standard 2PI method. 
To discuss the relative difficulty of the two calculations, we define $N_G$ and $N_M$ as the number of iterations required to obtain convergence of the self-energy and $M$ equations, respectively.

\vspace*{2mm}

\noindent {\it 2PI} (see section \ref{2PI-iterations-section})

\noindent (i) The zero temperature calculation requires $N_G\times N_M$ iterations. 

\noindent (ii) At finite temperatures the 2PI calculation takes $N_G + N_M$ iterations.  


\noindent {\it RG}

\noindent (i) At zero temperature the self-consistent equation for $\partial_\kappa\Sigma_\kappa$ must be solved $N_\kappa$ times. The vertex $M_0$ is obtained using the $G_0$ and relaxing the $M$ equation once. In addition, we must repeat the procedure multiple times to carry out the tuning process. 
The total number of relaxations at zero $T$ is $\left[(N_{G}\times N_\kappa)+N_M)\right]\times N_{\rm tune}$ where $N_{\rm tune}$ is the number of tuning steps that are needed. 

\noindent (ii) At finite temperature we need $(N_\kappa \times N_{G})+N_M$ relaxations.

In spite of the fact that it is slower, the RG method is interesting from a formal point of view, since it gives a different perspective on the theory of renormalization. 
The renormalization of the 2PI effective theory was discussed in section \ref{renorm-2PI-section}. The key points are 

\noindent (1) One must divide the counter-term contributions to the effective action into two pieces, which we have called $\Phi_{\Delta}$ and $\Phi_{\rm ct}$ (see equations (\ref{L-Delta}, \ref{L-ct})). The first is included, with the same form, at all loop orders. The second changes when the order of the approximation changes. 

\noindent (2) The different vertex counter-terms must be calculated by imposing renormalization conditions on the different 4-point functions in a precisely determined way, . 

\noindent Both of these points can be understood only by
performing a complicated diagrammatic analysis of the sub-divergences that are contained in the non-perturbative integral equations. 

The renormalization of higher order $n$PI theories is even more difficult to study. A
diagrammatic analysis seems prohibitively difficult, and  no other techniques have been previously available. 
The 2PI-RG method we have developed in this paper seems more straightforward to extend to higher order $n$PI theories. The number of bare parameters is fixed by the structure of the Lagrangian, and a complicated interdependent set of counter-terms is replaced by a hierarchy of flow equations that are straightforward to derive using the technique developed in this paper. Renormalization conditions can be enforced on higher order Bethe-Salpeter equations \cite{Russell2013}. The RG formalism is therefore a promising approach to the renormalization of higher order $n$PI theories. Work in this direction is in progress.


\newpage

\appendix


\section{Asymptotic limits of the solutions of the flow equations}
\label{appendixB}

The quantity in square brackets in (\ref{ass-prop}) is
\bea
\label{square1}
z = \big[\Sigma_\kappa(P)-\big(\Sigma_{\mu}(0)+P^2 \Sigma^\prime_{\mu}(0)\big)\big]\,.
\eea
It is easy to see that $z$ approaches zero when $\kappa \to\mu \gg P$ for any diagram in the expansion of the 2PI self-energy. 
As an example, we consider the sunset diagram which gives 
\bea
\Sigma_\mu(P)=
 -\frac{\lambda^2}{6}\int dQ\int dL \,G_\mu(Q)G_\mu(L)G_\mu(L+P+Q)\,.
\eea
If we rescale all momentum variables $Q=\hat Q\mu$, $L=\hat L\mu$, $P=\hat P\mu$ and $R_\mu(Q)=\mu^2 \hat R(\hat Q)$ and define $\hat G_{\mu}(\hat Q) = (\hat Q^2+\hat R(\hat Q)+m^2/\mu^2)^{-1}$ we obtain
\bea
\label{sunset}
\Sigma_\mu(P)= -\frac{\lambda^2}{6}\mu^2 \int d\hat Q\int d\hat L \hat G_\mu(\hat Q)\hat G_\mu(\hat L)\hat G_\mu(\hat L+\hat P+\hat Q)\,.
\eea
Assuming that $\hat R$ is a smooth function of its argument which respects rotational symmetry and expanding around $\hat P^2=0$, is it clear that $z$ goes to zero as $P^4/\mu^2$ for $\kappa \to\mu \gg P$.

The square bracket  in (\ref{lam4}) is
\bea
\label{lam4-extra}
z= \big[\Lambda^{\rm loop}_\kappa(P,Q)-\Lambda^{\rm loop}_\mu(0,0)\big]. 
\eea
To see that $z$ goes to zero in the limit $\kappa \to\mu \gg \{P,Q\}$ for any 2PI contribution to the kernel, we consider the example of the $t$-channel contribution to the kernel $\Lambda_\mu(P,Q)$. Using the dimensionless variables defined above we have
\bea
\Lambda_\mu(P,Q) = \frac{1}{2}\lambda^2\int dL G_\mu(P+Q+L)G_\mu(L) = \frac{1}{2}\lambda^2\int d\hat L \hat G_\mu(\hat P+\hat Q+\hat L)\hat G_\mu(\hat L)\,.
\eea
Equation (\ref{lam4-extra}) goes to zero as $P^2/\mu^2$ for $\kappa \to\mu \gg \{P,Q\}$.

\end{document}